\documentclass[12pt]{article}
\usepackage{geometry}
\usepackage{array}
\usepackage{longtable}
\usepackage{afterpage}
\usepackage{pdflscape}
\usepackage{arydshln}
\usepackage{multirow}
\usepackage{float}
\usepackage{url}
\usepackage{amsmath}
\usepackage{amsthm}
\usepackage{eufrak}
\usepackage{bbm}
\usepackage{graphicx}
\usepackage{esint}
\usepackage{setspace}
\usepackage{titlesec}
\usepackage{color}
\usepackage[acronym,nonumberlist]{glossaries}
\usepackage{natbib}

\titleformat{\section}{\large\bfseries}{\thesection}{1em}{}
\titleformat{\subsection}{\bfseries}{\thesubsection}{1em}{}
\titleformat{\subsubsection}{\itshape}{\thesubsubsection}{1em}{}

\DeclareMathAlphabet\mathbfcal{OMS}{cmsy}{b}{n}

\newcommand{\Keywords}[1]{\par\noindent
{\small{\em Keywords\/}: #1}}

\theoremstyle{definition}

\newacronym{ap}{AP}{aggregation property}
\newacronym{ap1}{AP1}{aggregation property}
\newacronym{ap2}{AP2}{aggregation property}
\newacronym{apn}{APN}{aggregation property}
\newacronym{apb}{APB}{aggregation property}
\newacronym{clr}{CONV}{cubed log return}
\newacronym{di}{DI}{discretisation-invariant}
\newacronym{hfrv}{HFRV}{high-frequency realised variance}
\newacronym{mtm}{MtM}{mark-to-market}
\newacronym{ntm}{NTM}{Neuberger's third moment}
\newacronym{lfrv}{LFRV}{low-frequency realised variance}
\newacronym{lv}{LV}{log variance}
\newacronym{otm}{OTM}{out-of-the-money}
\newacronym{pnl}{P\&L}{profit and loss}
\newacronym{qlr}{CONV}{fourth power of log return}
\newacronym{qv}{QV}{quadratic variation}
\newacronym{rfm}{RFM}{realised fourth moment}
\newacronym{rtm}{RTM}{realised third moment}
\newacronym{rv}{RV}{realised variance}
\newacronym{slr}{CONV}{squared log return}
\newacronym{snp}{S\&P 500}{Standard \& Poor's 500 Stock Market Index}
\newacronym{sdi}{SDI}{strike-discretisation-invariant}
\newacronym{vix}{VIX}{CBOE Volatility Index}

\begin{document}

\title{\Large The Aggregation Property\\ and its Applications to Realised Higher Moments}
\author{\large Carol Alexander and Johannes Rauch\footnote{School of Business, Management and Economics, University of Sussex, United Kingdom. Carol Alexander: c.alexander@sussex.ac.uk; Johannes Rauch: j.rauch@sussex.ac.uk.}}

\maketitle

\doublespacing

\thispagestyle{empty}

\begin{abstract}\normalsize
\noindent We develop a general multivariate aggregation property which encompasses the distinct versions of the property that were introduced by \cite{N12} and \cite{B14} independently. This way, we classify new types of model-free realised characteristics for which risk premia may be estimated without bias. We focus on the aggregation property for multivariate martingales and log martingales, and then define realised third and fourth moments which allow long-term higher-moment risk premia to be measured, efficiently and without bias, using high-frequency returns.\\
\end{abstract}

\clearpage
\glsresetall
\setcounter{page}{1}

\noindent 
The aggregation property in finance refers to the invariance of the expected value of a realised characteristic under  discretisation of the time interval over which the characteristic is measured. That is, the expected value remains the same irrespective of the partition used to monitor the realised characteristic. For example, if a forward price $F_t$ is a martingale under some measure, then the expected value (under that measure) of the sum of squared changes in the forward price until time $T$ is independent of the sampling frequency. In particular, the expected value is the same -- and equal to the expected value of $(F_T-F_0)^2$ -- whether the changes are monitored continuously, hourly, daily or weekly over the interval $[0,T]$. The partition of the interval doesn't even need to be a regular one.

The standard definition of realised variance -- an average squared log return -- does not satisfy the aggregation property. Therefore the variance risk premium, measured as the difference between the realised variance under the physical measure and the variance swap rate implied from option prices, is biased. By the same token, the theoretical fair value of a conventional variance swap can only be approximated, because the floating leg (realised variance) is computed as the average squared daily log return, whereas the theoretical value assumes the swap is continuously monitored. Consequently, market swap rates can deviate well beyond the no-arbitrage range, especially during crisis periods, which is when trading in volatility products increases. For example, during the financial crisis in 2008, variance swap rates for the \gls{snp} were frequently 5\% or more above the fair value determined by the \gls{vix} -- see \cite{AK12} and \cite{KS14}. The market for variance swaps (and their exchange-traded derivatives) is large, because they are excellent instruments for diversifying investment portfolios and transferring volatility risk.\footnote{They were introduced over-the-counter in the 1990's \citep{DDKZ99} and their futures, options, notes, funds and other derivatives are now being actively traded on most large exchanges, with demand stemming from their role as a diversifier, as a hedge, or purely for speculation, as illustrated by \cite{AKK15}. Currently, CBOE data show that \$3-\$6bn notional is traded daily on VIX futures contracts alone and on stock exchanges around the world even small investors can buy and sell over a hundred listed products linked to volatility futures. The most popular of these is Barclay's VXX note, with a market cap of around \$1 trillion.} As a result the literature on a variety of discretisation and model-dependent errors in variance swap rates is considerable.

If one re-defines realised variance so that it satisfies the aggregation property then there exists an exact, model-free fair-value variance swap rate under the minimal assumption of no arbitrage. Also, the same rate applies irrespective of the monitoring frequency of the floating leg. The expectation of the floating leg is path-independent, and even if investors differ in their views about jump risk in an incomplete market they will still agree on the fair-value swap rate. Furthermore, the fair-value swap rate can be expressed in terms of vanilla options written on the underlying, by using the replication theorem of \cite{CM01}.

Motivated by the search for more general variance characteristics which satisfy the aggregation property, \cite{N12} and \cite{B14} independently provide different definitions for this property. It is not possible to write one  in terms of the other, but our paper introduces a general aggregation property which encompasses both definitions, each as a different special case. We characterise a new class of realised pay-offs which satisfy the property, for multivariate martingales and  log martingales, including new definitions of higher moments of either price changes or log returns.

The \gls{apn} defined by \cite{N12}  is as follows: given an adapted stochastic process $\mathbf{x}_t$ on a standard filtered probability space and the real-valued function $g(\mathbf{x})$, the pair $\left(g;\mathbf{x}\right)$ satisfies the \gls{apn} iff
 \begin{equation}\label{eq:apn}\tag{APN}
 \mathbbm{E}_r\left[g\left(\mathbf{x}_T-\mathbf{x}_r\right)\right]=\mathbbm{E}_r\left[g\left(\mathbf{x}_s-\mathbf{x}_r\right)\right]+\mathbbm{E}_r\left[g\left(\mathbf{x}_T-\mathbf{x}_s\right)\right]\quad\forall\quad 0\le r\le s\le T,
\end{equation}
where $\mathbbm{E}_t$ denotes the expectation under some probability measure conditional on the filtration at time $t$. Applying the tower law of conditional expectations to \eqref{eq:apn} yields 
\begin{equation*}
\mathbbm{E}_0\left[\sum_{i=1}^Ng\left(\delta\mathbf{x}_i\right)\right]=\mathbbm{E}_0\left[g\left(\mathbf{x}_T-\mathbf{x}_0\right)\right],
\end{equation*}
for any partition $\left\{0=t_0<t_1<\ldots<t_N=T\right\}$, where $\delta\mathbf{x}_i=\mathbf{x}_{t_{i}}-\mathbf{x}_{t_{i-1}}$. That is, if \gls{apn} holds under a pricing measure, then the realised characteristic on the left has the same market price as the path-independent pay-off on the right.

When $\mathbf{x}=F$ is the forward price of a single tradeable asset the only functions which satisfy \eqref{eq:apn} are $g\left(\delta F\right)=\delta F$ and $g\left(\delta F\right)=\left(\delta F\right)^2$, where $\delta F$ denotes an increment in $F$. However, \cite{N12} focuses on two bivariate cases: the `arithmetic' case, $\mathbf{x}=\left(F, v\right)^\top$, where $v_t$ is the conditional variance of $F_T$; and the `geometric' case, $\mathbf{x}=\left(y, v^\phi\right)^\top$, where $y=\ln F$ denotes the log forward price and $v_t^\phi$ is some `generalised' variance process (defined later in this paper). This allows him to find new  second moments which, unlike the sum of squared log returns, satisfy \eqref{eq:apn}. He also finds one  third moment for which \gls{apn} holds,\footnote{See equation (5) p.3430 for the arithmetic case and Proposition 6, p.3435 for the geometric case.} and uses this moment to infer the skewness of long-term return distributions from observations on daily log returns.\footnote{\cite{KNS13} use the same realised moment to analyse the relationship between the (now unbiased) variance and skewness risk premiums in equity indices, finding that they are closely related.} He concludes by stating  ``[...] \textit{it would also be nice to be able to extend the analysis to higher-order moments. This would not be straightforward.}" 
Our generalised \gls{ap} yields an entire vector space of characteristics  which satisfy the \gls{ap}, including characteristics that allow higher-moment risk premia to be measured without the bias that arises from discretisation and jumps when standard moment definitions are employed.

\cite{B14} introduces an alternative version of the \gls{apb} which is based on the levels rather than the increments in a univariate process $x_t$ and a $\mathcal{C}^1$ function $h:\mathbbm{R}\times\mathbbm{R}\rightarrow\mathbbm{R}$. He states the property as:
 \begin{equation}\label{eq:apb}\tag{APB}
\mathbbm{E}_r\left[h\left(x_r, x_T\right)\right]=\mathbbm{E}_r\left[h\left(x_r, x_s\right)\right]+\mathbbm{E}_r\left[h\left(x_s, x_T\right)\right]\quad\forall\quad 0\le r\le s\le T.
\end{equation}
He shows that, if $x=F$ is a martingale (e.g. a forward price), the solutions to \gls{apb} are given by $h\left(x_r,x_s\right)=a\left(x_s\right)-a\left(x_r\right)+b\left(x_r\right)\left(x_s-x_r\right)$ for some real-valued functions $a$ and $b$. The $a$ part corresponds to the trivial solution, since $h\left(x_r, x_s\right)=a\left(x_s\right)-a\left(x_r\right)$ satisfies \gls{apb} for any process $x$. The $b$ part depends on the martingale assumption and disappears under expectation.\footnote{\cite{B14} further examines the special case where $b=-a^\prime$, see Equation 11 and Corollary 1. This restricted set of functions also appears in \cite{ST15a}, Equation 4, in the context of realised divergence.} In the univariate case \gls{apb} is more general than \gls{apn} in the sense that if the pair $\left(g;x\right)$ satisfies \eqref{eq:apn} then $h(x_r,x_s)=g(x_s-x_r)$ satisfies \eqref{eq:apb}. However, \cite{B14} leaves the more general case of a vector process for future research, and that is what we present in our paper.


In the following: Section \ref{sec:lit} briefly summarises the background literature on conventional variance swaps; Section \ref{sec:ap} reviews the results of \cite{N12} and \cite{B14} and defines our notation; Section \ref{sec:thr} presents our theoretical results on the general aggregation property, characterising an entire vector space of unbiased estimators for the associated risk premia, and then considers the efficiency of these estimators; Section \ref{sec:rhm} selects some aggregating realised characteristics which correspond to higher moments, focusing on new unbiased and efficient realised third and fourth moments for log returns, and Section \ref{sec:conc} concludes. All proofs are in the Appendix.

\section{Background on Variance Swaps}\label{sec:lit}\glsresetall

A conventional variance swap of maturity $T$ defines the \gls{rv} as the average squared daily log return on some underlying over the term of the swap. 
The calculation of a fair-value  swap rate proceeds under the assumptions that the pricing measure is unique,\footnote{In an arbitrage-free market, as in \cite{HK79}, expected pay-offs may be computed in a risk-neutral measure. In a complete market the risk-neutral measure for a representative investor corresponds to a unique market implied measure, see \cite{BL78}.} and: (a) monitoring of the floating leg happens continuously; (b) the forward price of the underlying follows a pure diffusion process; (c)  vanilla options on the underlying with the same maturity as the swap are traded at a continuum of strikes. Then a unique and exact fair-value swap rate -- which under assumption (a) becomes the expected \gls{qv} of the log price -- is derived from market prices of these  options. 

However, in the real world none of these assumptions hold. \cite{CW09} discuss the idealised case (a) where the \gls{rv} becomes the \gls{qv} of log returns. Then, assuming that the underlying follows a generic jump-diffusion process, they apply the replication theorem of \cite{CM01} to prove that 
$\mathbbm{E}\left[\mbox{QV}\right]=2\int_{\mathbbm{R}^+}k^{-2}{q}(k)dk+\iota,$
where $\mathbbm{E}$ denotes the expectation under the pricing measure and ${q}(k)$ denotes the price of a vanilla \gls{otm} option with strike $k$ and maturity $T$.\footnote{When $k\le F_0$ the option is a put and when $k>F_0$ the option is a call. This choice of separation strike is standard in the variance swap literature, e.g. in \cite{BKM03}.} When the underlying price follows a pure diffusion as in (b) the jump error $\iota$ is zero. Regarding assumption (c), in practice the integral must be computed numerically using the vanilla options that are actually traded. \cite{JT05} address the problems attendant to this assumption and derive upper bounds for the so-called `truncation error'. Also based on a finite number of traded strikes, \cite{DO14} derive model-free arbitrage bounds for continuously-monitored variance swap rates and claim that market rates are surprisingly close to the lower bound. 

A major source of error in the fair-value swap rate stems from assumption (a) because floating legs must be monitored in discrete time. This `discrete-monitoring' error may be written $\varepsilon=\mathbbm{E}\left[\mbox{RV}-\mbox{QV}\right]$.
Then, in the generic jump-diffusion setting of \cite{CW09}, the fair-value swap rate for the realised variance  may  be written
$\mathbbm{E}\left[\mbox{RV}\right]=2\int_{\mathbbm{R}^+}k^{-2}{q}(k)dk+\iota+\varepsilon.$
There is a large body of research on these pricing errors: \cite{CL09} prove that the discrete monitoring error $\varepsilon$ is related to the third moment of returns; \cite{JKLP13} investigate the convergence of the discretely-monitored swap rate to its continuously-monitored counterpart and derive bounds on $\varepsilon$ that get tighter as the monitoring frequency increases; \cite{BCM14} generalise these results and provide conditions for signing $\varepsilon$;  \cite{HK12} derive model-free  bounds for $\varepsilon$; \cite{BJ08b} derive fair-value swap rates for discretely-monitored variance swaps under various stochastic volatility diffusion and jump models, claiming that for most realistic contract specifications $\varepsilon$ is smaller than the error due to violation of assumption (b); \cite{BC14} extend their analysis to include a much wider variety of processes by considering the asymptotic expansion of $\varepsilon$. Finally, \cite{RT17} derive bounds for the jump error $\iota$ and demonstrate, via simulations and an empirical study, that price jumps induce a systematic negative bias which is particularly apparent when there are large downward jumps.

 However, cutting through this strand of research, both \cite{N12} and \cite{B14} provide a new class of generalised variance contracts for which exact replication of the floating leg is possible,  provided only that the underlying price follows a martingale. The replication strategy consists of a static portfolio of standard options and a dynamic trading strategy in the underlying asset. In these contracts the exposure to  variance can also vary over time in response to market conditions, e.g. to increase with the underlying  price level. And the key to defining these contracts is the aggregation property.

\section{The  Aggregation Property}\label{sec:ap}\glsresetall

Let $\mathbf{x}_t\in\mathbbm{R}^n$ for $t\in\left[0,T\right]$ denote a multivariate adapted process on a standard filtered probability space and consider the derivative process $\mathbf{u}_t=\mathbf{u}\left(t,\mathbf{x}_t\right)$ for a vector-valued function $\mathbf{u}:\left[0,T\right]\times\mathbbm{R}^n\rightarrow\mathbbm{R}^n$ as well as a real-valued function $f:\mathbbm{R}^n\times\mathbbm{R}^n\rightarrow\mathbbm{R}$. This setting allows the unification of the aggregation properties of \cite{N12} and \cite{B14} in the single definition as follows:
\begin{equation}\label{eq:ap}\tag{AP}
\mathbbm{E}_r\left[f\left(\mathbf{u}_r,\mathbf{u}_T\right)\right]=\mathbbm{E}_r\left[f\left(\mathbf{u}_r,\mathbf{u}_s\right)\right]+\mathbbm{E}_r\left[f\left(\mathbf{u}_s,\mathbf{u}_T\right)\right]\quad\forall\quad 0\le r\le s\le T.
\end{equation}
Our \gls{ap} is a joint condition on the pair $\left(f;\mathbf{u}\right)$. Making strong structural assumptions on one gives more flexibility to the other. For example, if $f\left(\mathbf{u}_r,\mathbf{u}_s\right)=a\left(\mathbf{u}_s\right)-a\left(\mathbf{u}_r\right)$ for some function $a:\mathbbm{R}^n\rightarrow\mathbbm{R}$, all processes $\mathbf{u}$ are a solution; and if $\mathbf{u}$ is constant, then all $f$ with $f\left(\mathbf{u},\mathbf{u}\right)=0$ are a solution. 

If \eqref{eq:ap} holds for $\left(f;\mathbf{u}\right)$, then by the tower law of expectations
\begin{equation}\label{eq:rc}\tag{E}
\mathbbm{E}_0\left[\sum_{i=1}^Nf\left(\mathbf{u}_{i-1},\mathbf{u}_i\right)\right]=\mathbbm{E}_0\left[f\left(\mathbf{u}_0,\mathbf{u}_T\right)\right],
\end{equation}
for all partitions $\left\{0=t_0<t_1<\ldots<t_N=T\right\}$, where we write $\mathbf{u}_i=\mathbf{u}_{t_i}$ for convenience. Following \cite{N12}, the interpretation of \eqref{eq:rc} depends on the measure: if \eqref{eq:ap} holds under the physical measure, then $\sum_{i=1}^Nf\left(\mathbf{u}_{i-1},\mathbf{u}_i\right)$ is an unbiased estimator of $\mathbbm{E}_0\left[f\left(\mathbf{u}_0,\mathbf{u}_T\right)\right]$ for any partition of $\left[0,T\right]$. If \eqref{eq:ap} holds under a pricing measure, then the fair price of a contingent claim that pays $\sum_{i=1}^Nf\left(\mathbf{u}_{i-1},\mathbf{u}_i\right)$ is the same as the price of a contingent claim that pays $f\left(\mathbf{u}_0,\mathbf{u}_T\right)$. Under the additional assumption that this contingent claim exists or can be synthesised from other claims, a fair price can be derived from the market.

Our \eqref{eq:ap} is more general than \eqref{eq:apb} in that we consider a multivariate (not necessarily martingale) derivative process, and more general than \eqref{eq:apn} in that our function is defined on the levels at the start and end of an interval in the partition rather than increments $\delta F$ over successive intervals. In fact, it would be possible to write \eqref{eq:apb} in terms of \eqref{eq:apn}, by doubling the size of the state space. That is, setting $\mathbf{x}=\left(F,y\right)^\top$, with $y=\ln F$, we may write
\begin{equation*}
g\left(\delta F,\delta y\right)=h\left(\frac{\delta F}{\mathrm{e}^{\delta y}-1},\frac{\delta F\mathrm{e}^{\delta y}}{\mathrm{e}^{\delta y}-1}\right), \quad \delta F, \delta y\ne 0.
\end{equation*}
However, the induced function $g$ becomes ill-defined as $\delta F, \delta y\rightarrow 0$ even when $h$ is a polynomial or some other well-behaved function.\footnote{We thank the associate editor and an anonymous referee for helpful comments in this regard. In fact, \cite{N12} doubles the size of the state space, but  the conditional variance process $v_t$ he includes in $\mathbf{x}$ 
	cannot be expressed in terms of $F_t$, unlike $y_t$. } Hence, \eqref{eq:apn} does not include \eqref{eq:apb}.

Clearly \eqref{eq:apn} is the special case of \eqref{eq:ap} where $f\left(\mathbf{u}_r,\mathbf{u}_s\right)=g\left(\mathbf{x}_s-\mathbf{x}_r\right)$. \cite{N12} characterises the solutions $\left(g;\mathbf{x}\right)$ to \eqref{eq:apn} when $\mathbf{x}$ is bivariate and the second component is a conditional expectation of the first component, whence the process has an implicit dependence structure. First \cite{N12} considers the arithmetic case $\mathbf{x}=\left(F,v\right)^\top$, where $v_t=\mathbbm{E}_t\left[\left(F_T-F_t\right)^2\right]$, and finds the solutions
\begin{center}
$g\left(\delta\mathbf{x}\right)=\left(c_1,c_2\right)^\top\delta\mathbf{x}+c_3\left(\delta F\right)^2+c_4\left\{\left(\delta F\right)^3+3\delta F\delta v\right\}$,
\end{center}
where $\delta\mathbf{x}$ denotes an increment in $\mathbf{x}$ and $c_i\in\mathbbm{R}$, $i\in\left\{1,\ldots,4\right\}$ are arbitrary real coefficients. The $\left(c_1,c_2\right)$ term satisfies \gls{apn} trivially, the $c_3$ term occurs provided $F$ is a martingale, since martingales have zero autocorrelation, and the $c_4$ term yields a characteristic which corresponds to a third moment, in the sense that
\begin{center}
$\mathbbm{E}_0\left[\sum_{i=1}^N\left\{\left(\delta F_i\right)^3+3\delta F_i\delta v_i\right\} \right]=\mathbbm{E}_0\left[\left(F_T-F_0\right)^3\right]$.
\end{center}
Then he considers the geometric case $\mathbf{x}=\left(y,v^\phi\right)^\top$, where $y=\ln F$ and $v_t^\phi=\mathbbm{E}_t\left[\phi\left(y_T-y_t\right)\right]$ denotes a generalised variance process, i.e. any process such that $\lim_{\delta y\rightarrow 0}\phi\left(\delta y\right)/\left(\delta y\right)^2=1$. In this case he shows that the solutions to \eqref{eq:apn} are given by
\begin{center}
$g\left(\delta\mathbf{x}\right)=\left(c_5,c_6\right)^\top\delta\mathbf{x}+c_7\left(\mathrm{e}^{\delta y}-1\right)+c_8\left(\delta v^\phi-2\delta y\right)^2+c_9\left(\delta v^\phi+2\delta y\right)\mathrm{e}^{\delta y},$
\end{center}
where $c_i\in\mathbbm{R}$, $i\in\left\{5,\ldots,9\right\}$, and at least one of $c_8$ and $c_9$ must be zero. The $\left(c_5,c_6\right)$ term satisfies \gls{apn} trivially and the $c_7$ term has zero expectation, provided $F$ is a martingale. Most interesting are the $c_8$ and $c_9$ terms: when
$c_8\,{\ne}\,0$ the generalised variance process is called the `log variance' process, denoted $v_t^{\lambda}=\mathbbm{E}_t\left[\lambda\left(y_T-y_t\right)\right]$, where $\lambda\left(\delta y\right)=2\left(\mathrm{e}^{\delta y}-1-\delta y\right)$. When $c_9\,{\ne}\,0$ the generalised variance process is called the `entropy variance' process, denoted  $v_t^{\eta}=\mathbbm{E}_t\left[\eta\left(y_T-y_t\right)\right]$, where $\eta\left(\delta y\right)=2\left(\delta y\mathrm{e}^{\delta y}-\mathrm{e}^{\delta y}+1\right)$; and when $c_8=c_9=0$ then $v^\phi$ can be any generalised variance process.

Within this geometric set of solutions to \eqref{eq:apn}, \cite{N12} focuses on one particular realised variance, i.e. the \gls{lv}, for which $\left(c_5,c_6,c_7,c_8,c_9\right)=\left(-2,0,2,0,0\right)$, so that $g\left(\delta y\right)=2\left(\mathrm{e}^{\delta y}-1-\delta y\right)=\lambda\left(\delta y\right)$.\footnote{Note that $2\left(\delta y\mathrm{e}^{\delta y}-\mathrm{e}^{\delta y}+1\right)$, i.e. the realised variance corresponding to the entropy variance process, does not satisfy \eqref{eq:apn}. That is, there is no choice of $\left(c_5,c_6,c_7,c_8,c_9\right)$ which yields $g\left(\delta y\right)=\eta\left(\delta y\right)$.} He finds only one higher-order moment, corresponding to $\left(c_5,c_6,c_7,c_8,c_9\right)=\left(6,-3,-12,0,3\right)$, given by
\begin{center}
$g\left(\delta\mathbf{x}\right)= \rho(\delta\mathbf{x})
+\tau\left(\delta y\right),$\\ where 
 $\rho(\delta\mathbf{x})=3\delta v^\eta\left(\mathrm e^{\delta y}-1\right)$ and $\tau\left(\delta y\right)=6\left(\delta y\mathrm e^{\delta y}-2\mathrm e^{\delta y}+\delta y+2\right)$.\\
\end{center}
To find the long-term moment corresponding to this choice of $g$, set $\delta\mathbf{x}= \mathbf{x}_T - \mathbf{x}_0$ and take expectations. Since $\mathbbm{E}_0\left[\rho\left(\mathbf{x}_T-\mathbf{x}_0\right)\right]=0$ when $F$ is a martingale, we have $\mathbbm{E}_0\left[g\left(\mathbf{x}_T-\mathbf{x}_0\right)\right]=\mathbbm{E}_0\left[\tau\left(y_T-y_0\right)\right]$. Unfortunately, even though $\lim_{\delta y\rightarrow 0}\tau\left(\delta y\right)/\left(\delta y\right)^3=1$,  the implied characteristic does not capture a third moment because it will be dominated by terms in $\tau$ with order greater than 3 when  $y_T-y_0$ is sufficiently large. This motivates our search for a new third-moment solution to \eqref{eq:ap} for which the long-term moment corresponds to a moment of order exactly 3.\footnote{The fact that $\tau\left(\delta y\right)$ is O$\left(\delta y^3\right)$ is useful for measurement based on high-frequency data. But $\rho(\delta\mathbf{x})$ is O$\left(\delta v^\eta\delta y\right)$, so $g\left(\delta\mathbf{x}\right)$ is not a pure cubic exposure because it includes an additional price-variance covariance exposure. 
	See \cite{N12} for further discussion on this point. } 

\section{Theoretical Results}\label{sec:thr}\glsresetall

Our first result characterises the pairs $\left(f; \mathbf{u}\right)$, with $f:\mathbbm{R}^n\times\mathbbm{R}^n\rightarrow\mathbbm{R}$, which satisfy \eqref{eq:ap} as solutions to two second-order partial differential equations, one for the derivative processes $\mathbf{u}=\left(u_{1},\ldots,u_{n}\right)^\top$  w.r.t. the underlying processes $\mathbf{x}=\left(x_{1},\ldots,x_{n}\right)^\top$ and the other for the real-valued function $f$ w.r.t. $\mathbf{u}$. For this we need to assume that $f$ and $\mathbf{u}$  are twice differentiable, so that the following quantities exist:  For $i\in\left\{1,\ldots,n\right\}$ let $\boldsymbol{\vartheta}_t=\left(\vartheta_{1t},\ldots,\vartheta_{nt}\right)^\top$ where  $\vartheta_{it}=\vartheta_i\left(t,\mathbf{x}_t\right)\in\mathbbm{R}$ denote the time derivatives of $\mathbf{u}$, and let  $\boldsymbol\delta_{it}=\boldsymbol\delta_i\left(t,\mathbf{x}_t\right)\in\mathbbm{R}^n$ and $\boldsymbol\Gamma_{it}=\boldsymbol\Gamma_i\left(t,\mathbf{x}_t\right)\in\mathbbm{R}^{n\times n}$ denote the first and second partial derivatives of $\mathbf{u}_{it}$ w.r.t. the components of $\mathbf{x}_t$. For the function $f$ we denote the Jacobian vector of first partial derivatives w.r.t. the components of the second input vector by $\mathbf{J}=\left(J_1,\ldots, J_n\right)^\top\in\mathbbm{R}^n$ and write $\mathbf{H}\in\mathbbm{R}^{n\times n}$ for the Hessian matrix of second partial derivatives. With these definitions we can now establish \textit{necessary} conditions for $\left(f; \mathbf{u}\right)$ to satisfy the \eqref{eq:ap}, by considering a particular process  for $\mathbf{x}_t$ (a multivariate diffusion with a particular drift and covariance) and then deriving conditions for the \gls{ap} to hold.\\

\noindent {\bfseries Theorem 1:} Assume that  $\mathbf{x}$ follows a diffusion process with dynamics $d\mathbf{x}_t=\boldsymbol\mu_tdt+\boldsymbol\Sigma_td\mathbf{w}_t$, where $\boldsymbol\Sigma_t=\boldsymbol\Sigma\left(t,\mathbf{x}_t\right)\in\mathbbm{R}^{n\times n}$, $\boldsymbol\mu_t=\boldsymbol\mu\left(t,\mathbf{x}_t\right)\in\mathbbm{R}^n$ and $\mathbf{w}_t\in\mathbbm{R}^n$ is a standard multivariate Wiener process, for $t \in [0,T]$. If the first derivatives of $\mathbf{u}$ w.r.t. $\mathbf{x}$, at time $t$, viz. $\boldsymbol\Delta_t=\left(\boldsymbol\delta_{1t},\ldots,\boldsymbol\delta_{nt}\right)^\top$ form an $n \times n$ invertible matrix 
then for every $\left(f;\mathbf{u}\right)$ such that $f\left(\mathbf{u},\mathbf{u}\right)=0 \,\, \forall \,\, \mathbf{u}\in\mathbbm{R}^n$, there exists a function $a\in\mathcal{C}^2$ such that:
\begin{equation}\label{eq:pde1}\tag{PDE\ I}
\mathbf{H}\left(\mathbf{u}_r,\mathbf{u}_s\right)-\mathbf{H}^a\left(\mathbf{u}_s\right)=\sum_{i=1}^n\left\{J_i\left(\mathbf{u}_r,\mathbf{u}_s\right)-J_i^a\left(\mathbf{u}_s\right)\right\}\mathbf{C}_{is},
\end{equation}
for any $0\le r\le s\le T$, where $\mathbf{J}^a=\left(J_1^a,\ldots, J_n^a\right)^\top\in\mathbbm{R}^n$ and $\mathbf{H}^a\in\mathbbm{R}^{n\times n}$ are the Jacobian and Hessian of $a$ and $\mathbf{C}_{it}=\mathbf{C}_i\left(\mathbf{u}_t\right)\in\mathbbm{R}^{n\times n}$ are symmetric matrices, for $i=1,\ldots,n$. Moreover:
\begin{equation}\label{eq:pde2}\tag{PDE\ II}
	2\left(\boldsymbol\vartheta_{t}+\boldsymbol\Delta_{t}\boldsymbol\mu_t\right)+\left(\text{tr}\left\{\boldsymbol\Sigma_t^\top\mathbf{M}_{1t}\boldsymbol\Sigma_t\right\},\ldots,\text{tr}\left\{\boldsymbol\Sigma_t^\top\mathbf{M}_{nt}\boldsymbol\Sigma_t\right\}\right)^\top=\mathbf{0},
\end{equation}
where $\mathbf{M}_{it}=\boldsymbol\Gamma_{it}+ \boldsymbol\Delta_t^\top\mathbf{C}_{it}\boldsymbol\Delta_t$, for all $t \in [0,T]$.\\

\noindent Having derived the necessary conditions \eqref{eq:pde1} and \eqref{eq:pde2} we now provide the solutions to \eqref{eq:ap} for symmetric matrices $\mathbf{C}_{it}$ and derive closed-form expressions for $\left(f;\mathbf{u}\right)$ in two special cases, viz. Corollary 2, when all $\mathbf{C}_{it}$ are constant -- so that each process in $\mathbf{u}$ follows a log martingale -- and Corollary 3 when all $\mathbf{C}_{it}$ are zero, in which case $\mathbf{u}$ contains only martingales. In each case the definition of the process $\mathbf{u}$ is otherwise model-free. Note that the underlying process $\mathbf{x}$ does not need to be a diffusion. This assumption was made in Theorem 1 merely to find necessary conditions for the solutions to \eqref{eq:ap}. Once a general form of solution is found we can verify that it satisfies our \gls{ap} for any underlying process simply by substitution in \eqref{eq:ap}. But first we need the following:\\

\noindent {\bfseries Corollary 1:} According to the multivariate Feynman-Kac formula with boundary condition $\mathbf{u}\left(T,\mathbf{x}_T\right)=\boldsymbol\psi\left(\mathbf{x}_T\right)$ the solution to \eqref{eq:pde2} is given by
\begin{center}
	$\mathbf{u}\left(t,\mathbf{x}_t\right)=\mathbbm{E}_t\left[\boldsymbol\psi\left(\mathbf{x}_T\right)+\tfrac{1}{2}\int_t^T\left(\text{tr}\left\{\boldsymbol\Sigma_\tau^\top\boldsymbol\Delta_\tau^\top\mathbf{C}_{1\tau}\boldsymbol\Delta_\tau\boldsymbol\Sigma_\tau\right\},\ldots,\text{tr}\left\{\boldsymbol\Sigma_\tau^\top\boldsymbol\Delta_\tau^\top\mathbf{C}_{n\tau}\boldsymbol\Delta_\tau\boldsymbol\Sigma_\tau\right\}\right)^\top d\tau\right].$
\end{center}
Hence, the derivative process $u_i$ follows a martingale if and only if $\mathbf{C}_{it}=\mathbf{0}$.\\

\noindent {\bfseries Theorem 2:} Suppose that $\mathbf{C}_{it}=\mathbf{Q}\mathbf{D}_{it}\mathbf{Q}^\top$, where $\mathbf{Q}$ is orthogonal and $\mathbf{D}_{it}$ are diagonal matrices, and that $\sum_{i=1}^n\mathbf{C}_{it}=\left(\mathbf{C}_{1t}\mathbf{1},\ldots,\mathbf{C}_{nt}\mathbf{1}\right)^\top$. 
Then the solution to \eqref{eq:pde1} is given by
\begin{equation*}
	f\left(\mathbf{u}_r,\mathbf{u}_s\right)=a\left(\mathbf{u}_s\right)-a\left(\mathbf{u}_r\right)+\mathbf{b}\left(\mathbf{u}_r\right)^\top\left\{\mathbf{m}\left(\mathbf{u}_s\right)-\mathbf{m}\left(\mathbf{u}_r\right)\right\},
\end{equation*}
where $a:\mathbbm{R}^n\rightarrow\mathbbm{R}\in\mathcal{C}^2$ with $a\left(\mathbf{0}\right)=0$, $\mathbf{b}:\mathbbm{R}^n\rightarrow\mathbbm{R}^n\in\mathcal{C}^1$, and
\begin{equation*}
\mathbf{m}\left(\mathbf{u}_t\right)=\int_0^{\mathbf{u}_t}\exp\left\{\sum_{i=1}^n\int_0^{u_i}\mathbf{C}_i\left(\mathbf{\tilde{u}}\right)d\tilde{u}_i\right\}d\mathbf{u},
\end{equation*}
is a multivariate martingale.\\

\noindent Again the $a$ term satisfies the \eqref{eq:ap} trivially, and the $\mathbf{b}$ term has zero expectation. If we do not impose $a\in\mathcal{C}^2$ and $\mathbf{b}\in\mathcal{C}^1$ there may be other functions $f$ which satisfy the \gls{ap}. 

The following Corollary characterises all solutions to \eqref{eq:pde1} and \eqref{eq:pde2} in the case that $\mathbf{u}$ follows a multivariate log martingale.\\

\noindent {\bfseries Corollary 2:} Suppose that $\mathbf{C}_{it}=\mathbf{C}_i$ are constant and that $\sum_{i=1}^n\mathbf{C}_i$ is invertible. Then
\begin{equation*}
\mathbf{m}\left(\mathbf{u}_t\right)=\left(\sum_{i=1}^n\mathbf{C}_i\right)^{-1}\left(\exp\left\{\sum_{i=1}^n\mathbf{C}_iu_{it}\right\}-\mathbf{I}\right)\mathbf{1},
\end{equation*}
and
\begin{equation*}
\mathbf{u}\left(t,\mathbf{x}_t\right)=\left(\sum_{i=1}^n\mathbf{C}_i\right)^{-1}\ln\left(\mathbbm{E}_t\left[\exp\left\{\sum_{i=1}^n\mathbf{C}_i\psi_i\left(\mathbf{x}_T\right)\right\}\right]\right)\mathbf{1},
\end{equation*}
the log of a martingale satisfying the boundary condition $\mathbf{u}_T=\boldsymbol\psi\left(\mathbf{x}_T\right)$.\\

\noindent The martingale case in the next Corollary corresponds to the limit as $\mathbf{C}_i\rightarrow\mathbf{0}$, $i=1,\ldots,n$ of the log martingale case in Corollary 2. It is straightforward to verify that the martingale condition \eqref{eq:msoln} below characterises all pairs $(f;\mathbf{u})$ which satisfy \eqref{eq:ap} when $\mathbf{u}$ is a martingale, for \textit{any}  $\mathcal{C}^2$ function $a$ and \textit{any} set of $\mathcal{C}^1$ functions in $\mathbf{b}$.\\ 

\noindent {\bfseries Corollary 3:} Suppose that $\mathbf{C}_i=\mathbf{0}$ for all $i\in\left\{1,\ldots,n\right\}$. Then the solution to \eqref{eq:pde1} is
\begin{equation}\label{eq:msoln}\tag{M}
f\left(\mathbf{u}_r,\mathbf{u}_s\right)=a\left(\mathbf{u}_s\right)-a\left(\mathbf{u}_r\right)+\mathbf{b}\left(\mathbf{u}_r\right)^\top\left(\mathbf{u}_s-\mathbf{u}_r\right),
\end{equation}
where we assume w.l.o.g. that $\mathbf{J}^a\left(\mathbf{0}\right)=\mathbf{0}$. Furthermore, the solution to \eqref{eq:pde2} is  $\mathbf{u}\left(t,\mathbf{x}_t\right)=\mathbbm{E}_t\left[\boldsymbol\psi\left(\mathbf{x}_T\right)\right]$.\\

\noindent Note that the univariate martingale case corresponds to the solutions described in equation (12) of \cite{B14} to the \gls{apb}.

 The realised characteristic based on $\left(f;\mathbf{u}\right)$, i.e.
\begin{equation*}
\sum_{i=1}^Nf\left(\mathbf{u}_{i-1},\mathbf{u}_i\right)=a\left(\mathbf{u}_T\right)-a\left(\mathbf{u}_0\right)+\sum_{i=1}^N\mathbf{b}\left(\mathbf{u}_{i-1}\right)^\top\left(\mathbf{u}_i-\mathbf{u}_{i-1}\right),
\end{equation*}
is an unbiased estimator for the implied characteristic $\mathbbm{E}_0\left[f\left(\mathbf{u}_0,\mathbf{u}_T\right)\right]$ which, by virtue of $\mathbf{u}$ being a martingale, may also be written $\mathbbm{E}_0\left[a\left(\mathbf{u}_T\right)\right]-a\left(\mathbf{u}_0\right)$. Note that $\mathbf{u}_i=\mathbf{u}_{t_i}$. Thus, the implied characteristic is defined by $a$ alone. In the next section we shall consider some particular choices for $a$ which correspond to higher moments.

While $a$ must be fixed at time 0, $\mathbf{b}$  can change dynamically over time. Moreover, $\mathbf{b}$ determines the conditional variance of the estimator along the partition, because
\begin{eqnarray*}
&&\mathbbm{E}_t\left[\left\{\sum_{i=1}^Nf\left(\mathbf{u}_{i-1},\mathbf{u}_i\right)-\mathbbm{E}_t\left[\sum_{i=1}^Nf\left(\mathbf{u}_{i-1},\mathbf{u}_i\right)\right]\right\}^2\right]\\
&=&\mathbbm{E}_t\left[\left\{a\left(\mathbf{u}_T\right)-\mathbbm{E}_t\left[a\left(\mathbf{u}_T\right)\right]+\sum_{t_i>t}\mathbf{b}\left(\mathbf{u}_{i-1}\right)^\top\left(\mathbf{u}_i-\mathbf{u}_{i-1}\right)\right\}^2\right],
\end{eqnarray*}
for $t\in\left\{t_0,\ldots,t_N\right\}$. Next we propose an optimal choice for $\mathbf{b}$ in the sense that it yields a conditionally efficient estimator, i.e. an estimator with minimum conditional variance:\\

\noindent {\bfseries Theorem 3:} Given a martingale process $\mathbf{u}$ and some function $a \in \mathcal{C}^2$ which specifies an implied characteristic for this process, the  conditionally efficient estimator for the implied characteristic $\mathbbm{E}_0\left[f\left(\mathbf{u}_0,\mathbf{u}_T\right)\right]$ has a first-order approximation given by $\mathbf{b}^\star\left(\mathbf{u}\right)=-\mathbf{J}^a\left(\mathbf{u}\right)$. 
 If the process is a diffusion and monitoring happens continuously, the approximation is exact.\\

\noindent Theorem 3 gives a first-order approximation which becomes more exact as the monitoring frequency of the realised characteristic increases. The approximation error is not as detrimental to accurate measurement of a risk premium as violations of the \gls{ap}. On a large sample small violations of \eqref{eq:ap} can result in a large cumulative bias, whereas finding efficiency only to a first-order approximation may be acceptable on a large sample with frequent monitoring.

\section{Examples of Aggregating Realised Higher Moments}\label{sec:rhm}

First we show that the solution \eqref{eq:msoln} encompasses the realised variances  and  third moment found by  \cite{N12}.\footnote{Expressing Bondarenko's generalised and power-price weighted variance contracts (p.88) in our notation is a straight-forward task, since his set of solutions corresponds to our univariate martingale case.} To this end we first re-write Neuberger's geometric set of solutions in an unrestricted fashion by setting $\mathbf{x}=\left(y,v^\lambda,v^\eta\right)^\top$ and writing
\begin{equation*}
g\left(\delta\mathbf{x}\right)=\left(c_5,c_6^\lambda,c_6^\eta\right)^\top\delta\mathbf{x}+c_7\left(\mathrm{e}^{\delta y}-1\right)+c_8\left(\delta v^\lambda-2\delta y\right)^2+c_9\left(\delta v^\eta+2\delta y\right)\mathrm{e}^{\delta y}.
\end{equation*}
Next we define the log contract and entropy contracts, viz. $Y_t=\mathbbm{E}_t\left[y_T\right]$ and  $Z_t=\mathbbm{E}_t\left[F_Ty_T\right]$, respectively. Note that $v^\lambda=2\left(y-Y\right)$ and $v^\eta=2\left(\tfrac{Z}{F}-y\right)$.\\

\noindent {\bfseries Corollary 4:} Let $\mathbf{u}=\left(F,Y,Z\right)^\top$ and set 
$a\left(\mathbf{u}\right)=\left(c_5+2c_6^\lambda-2c_6^\eta,4c_8,2c_6^\eta\right)\left(\ln F,Y^2,\tfrac{Z}{F}\right)^\top$
and 
$\mathbf{b}\left(\mathbf{u}\right)=\left(\tfrac{c_7}{F}-\tfrac{2c_9Z}{F^2},-2c_6^\lambda-8c_8Y,\tfrac{2c_9}{F}\right)^\top.$ 
Then 
\begin{equation*}
f\left(\mathbf{u}_r,\mathbf{u}_s\right)=a\left(\mathbf{u}_s\right)-a\left(\mathbf{u}_r\right)+\mathbf{b}\left(\mathbf{u}_r\right)^\top\left(\mathbf{u}_s-\mathbf{u}_r\right)=g\left(\mathbf{x}_s-\mathbf{x}_r\right),
\end{equation*}
and hence the solution to \eqref{eq:apn} can be expressed in terms of the solution to \eqref{eq:ap}.\\

\noindent For example, setting $\left(c_5,c_6^\lambda,c_6^\eta,c_7,c_8,c_9\right)=\left(-2,0,0,2,0,0\right)$ so that $a\left(\mathbf{u}\right)=-2\ln F$ and $\mathbf{b}\left(\mathbf{u}\right)=\left(\tfrac{2}{F},0,0\right)^\top$, we have
$f\left(\mathbf{u}_r,\mathbf{u}_s\right)=\lambda\left(y_s-y_r\right)$, 
which corresponds to Neuberger's \gls{lv}. Similarly, setting $\left(c_5,c_6^\lambda,c_6^\eta,c_7,c_8,c_9\right)=\left(6,0,-3,-12,0,3\right)$ yields $a\left(\mathbf{u}\right)=12\ln F-6\tfrac{Z}{F}$ and $\mathbf{b}\left(\mathbf{u}\right)=\left(\tfrac{-12}{F}-\tfrac{6Z}{F^2},0,\tfrac{6}{F}\right)^\top$, and then we have $f\left(\mathbf{u}_r,\mathbf{u}_s\right)=\rho\left(\mathbf{x}_s-\mathbf{x}_r\right)+\tau\left(y_s-y_r\right)$, 
which corresponds to \gls{ntm}. In both examples $\mathbf{b}\left(\mathbf{u}\right)=\mathbf{b}^\star\left(\mathbf{u}\right)$, so both NTM and \gls{lv} are approximately efficient. However, 
in the empirical work of \cite{N12} and \cite{KNS13} the vector $\mathbf{b}$ is fixed monthly rather than rebalanced at the daily monitoring frequency, which makes the estimator less efficient.\\

\noindent 
Next we suppose that $\mathbf{u}$ contains a series of power log contracts $P_t^{(i)}=\mathbbm{E}_t\left[y_T^i\right]$, $i\ge 0$ on a single underlying $F$. According to the replication theorem of \cite{CM01}, for $i\ge 1$ this conditional expectation can be expressed in terms of vanilla \gls{otm} options as:
\begin{equation*}\label{eq:powerlogcontracts}
P_t^{(i)}=y_t^i+\int_{\mathbbm{R}^+}\gamma_i(k)q_t(k)dk,
\end{equation*}
where $\gamma_i(k):=i(\ln k)^{i-2}k^{-2}\left[i-1-\ln k\right]$ and $q_t(k)$ denotes the time-$t$  price of a vanilla \gls{otm} option with strike $k$ and maturity $T$. In particular, for $i=1$, this yields the replication portfolio $Y_t=y_t-\int_{\mathbbm{R}^+}k^{-2}q_t(k)dk$ of the log contract. 
So, let $\mathbf{u}=\left(Y,P^{(2)}\ldots,P^{(n)}\right)^\top$ 
and consider the specification
\begin{equation}\label{eq:a}\tag{a}
a\left(\mathbf{u}\right)=n\left(-Y\right)^{n+1}-\sum_{i=2}^{n}\tbinom{n+1}{i}P^{(i)}\left(-Y\right)^{n+1-i}.
\end{equation}
Note that $a\left(\mathbf{u}_T\right)=y_T^{n+1}$, since $P_T^{(i)}=y_T^i$ and $\sum_{i=0}^{n+1}\tbinom{n+1}{i}\left(-1\right)^{n+1-i}=0$. Then, according to Corollary 3, and using that $P^{(0)}=1$ as well as $P^{(1)}=Y$, the implied characteristic equals
\begin{equation}\label{eq:cm}\tag{CM}
\mathbbm{E}_0\left[f\left(\mathbf{u}_0,\mathbf{u}_T\right)\right]=\mathbbm{E}_0\left[a\left(\mathbf{u}_T\right)\right]-a\left(\mathbf{u}_0\right)=\sum_{i=0}^{n+1}\tbinom{n+1}{i}P_0^{(i)}\left(-Y_0\right)^{n+1-i}=\mathbbm{E}_0\left[\left(y_T-Y_0\right)^{n+1}\right]
\end{equation}
which is the ${n+1}^{st}$ central moment of the log return distribution.

 \begin{figure}[h!]
 	\begin{centering}
 		\includegraphics[width=0.49\linewidth,trim=0cm 0cm 0cm 0cm,clip=true]{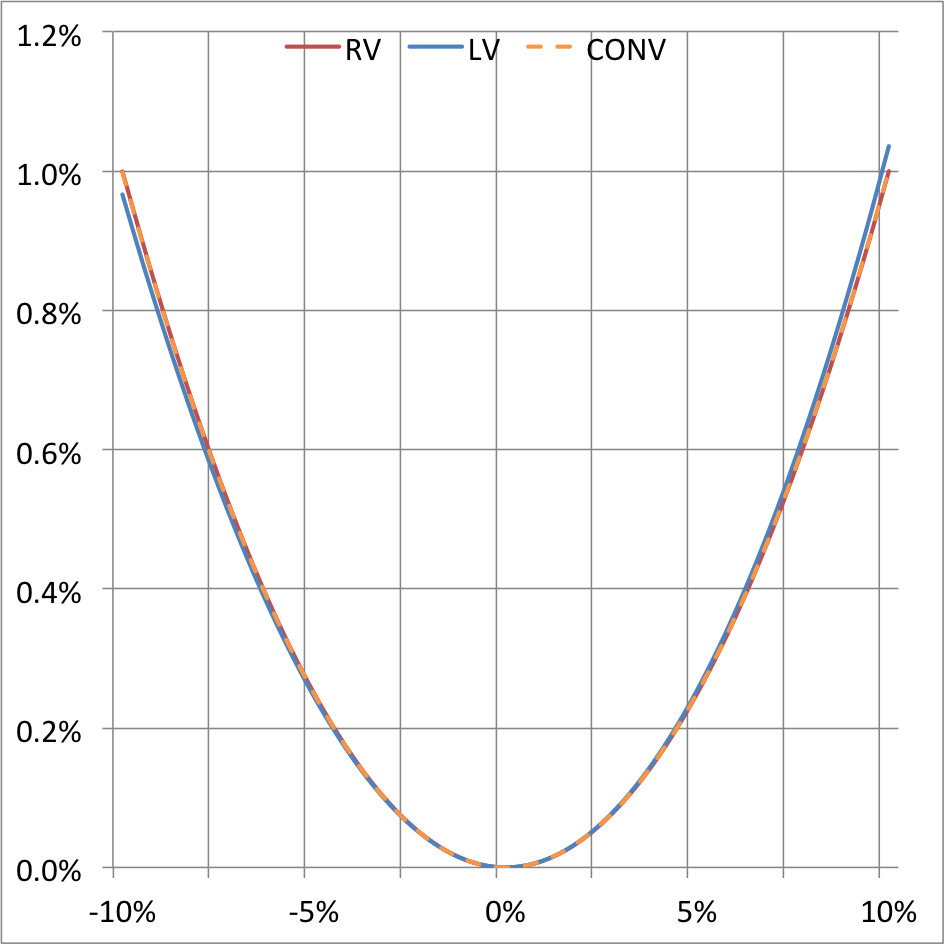}
 		\includegraphics[width=0.49\linewidth,trim=0cm 0cm 0cm 0cm,clip=true]{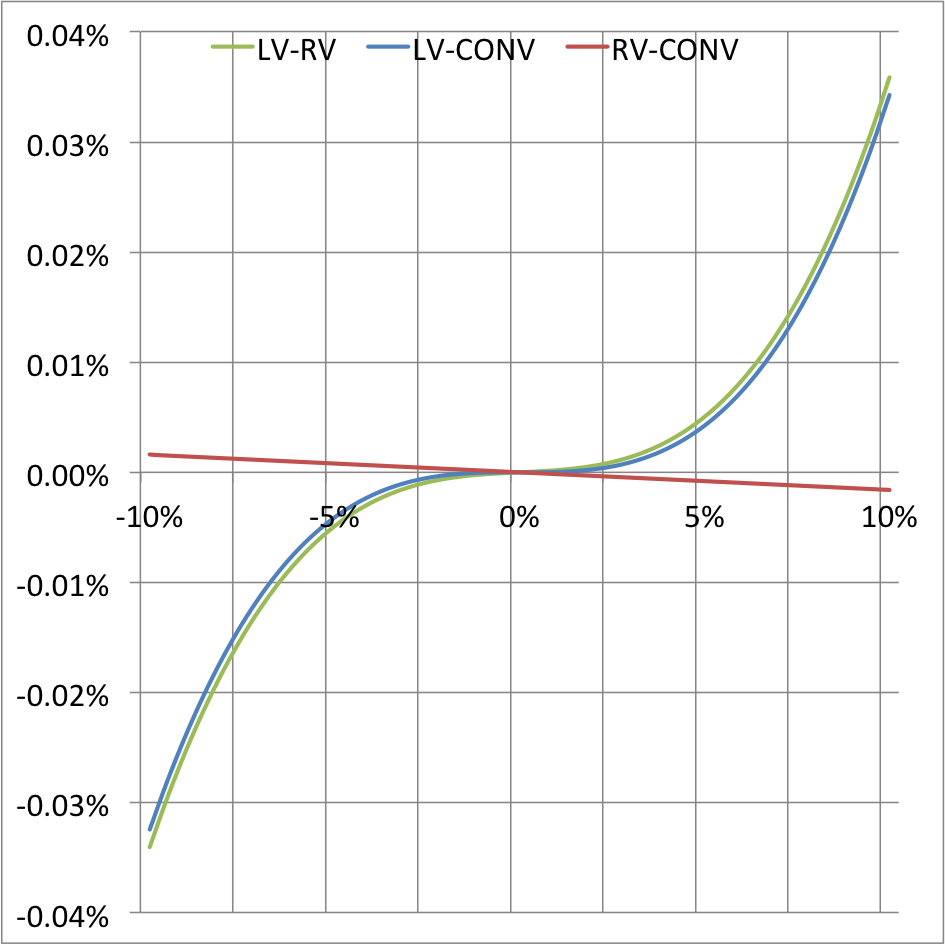}
 		\par\end{centering}
 	{\caption[Realised variance characteristics]{\small Comparison of our efficient \gls{rv}, Neuberger's \gls{lv} as well as the conventional \acrfull{slr} for $s-r=1/250$ (daily monitoring) and $\sigma=20\%$ (implied volatility). The $x$-axis corresponds to $Y_s-Y_r$, the change in price of the log contract. Note that $y_s-y_r=Y_s-Y_r+\tfrac{\sigma^2}{2}\left(s-r\right)$.}\label{fig:rvcharacteristics}}
\end{figure}

For $n=1$ we have $a\left(\mathbf{u}\right)=Y^2$ and, using Theorem 3, $\mathbf{b}^\star\left(\mathbf{u}\right)=-2Y$, so that
$f\left(\mathbf{u}_r,\mathbf{u}_s\right)
=\left(Y_s-Y_r\right)^2$, the squared change in price of the log contract.  This characteristic corresponds to the $c_8$ term in Neuberger's geometric set of solutions and it can be replicated by holding a squared log contract and shorting $2Y_r$ log contracts from time $r$ to time $s$. Here, and in the following, $s-r=1/250$ for daily monitoring.

Note that the calculation of $Y_t$ from option prices is subject to a numerical integration error, whereas the \gls{lv} may be derived from direct observation of the underlying price alone. For this reason, the LV is preferable to the squared change in price of the log contract for the unbiased (and efficient) estimation of a variance risk premium. However, there is very little difference between them in Figure \ref{fig:rvcharacteristics}. Here we compare the new RV estimator with existing definitions of realised variance, under the assumption that vanilla options with a continuum of strikes can be traded and therefore the log contract can be synthesised, using the replication theorem of \cite{CM01}. Note that this illustration is otherwise model-free, e.g. it allows for stochastic volatility or jumps in the underlying price process.

For $n\ge 2$ we apply Theorem 3 to derive the first-order efficient estimator for the  ${n+1}^{st}$ central moment of the log return distribution as: 
\begin{equation}\label{eq:b}\tag{b}
\mathbf{b}^\star\left(\mathbf{u}\right)=\left(n\left(n+1\right)\left(-Y\right)^n-\sum_{i=2}^{n}\tbinom{n+1}{i}\left(n+1-i\right)P^{(i)}\left(-Y\right)^{n-i},\ldots,\left(n+1\right)\left(-Y\right)\right)^\top.
\end{equation}
Now we use this to define a new realised third moment that does not suffer the same problem as NTM. More precisely, we seek an aggregating third moment for which the implied characteristic corresponds \textit{exactly} to the third central moment of the log return distribution. Setting $n=2$ in \eqref{eq:a} and \eqref{eq:b} yields $a\left(\mathbf{u}\right)=-2Y^3+3P^{(2)}Y$ as well as $\mathbf{b}^\star\left(\mathbf{u}\right)=\left(6Y^2-3P^{(2)},-3Y\right)^\top$, and therefore
\begin{equation}\label{eq:rtm}\tag{RTM}
f\left(\mathbf{u}_r,\mathbf{u}_s\right)
=\left(Y_s-Y_r\right)^3+3\left(v_s^{(2)}-v_r^{(2)}\right)\left(Y_s-Y_r\right),
\end{equation}
which is similar to the $c_4$ term in Neuberger's arithmetic set of solutions, but has the log contract rather than the forward price as the underlying. The pay-off can be replicated by holding a cubed log contract as well as $6Y_r^2-3P_r^{(2)}$ log contracts and shorting $3Y_r$ squared log contracts from time $r$ to time $s$.

\begin{figure}[h!]
	\begin{centering}
		\includegraphics[width=0.49\linewidth,trim=0cm 0cm 0cm 0cm,clip=true]{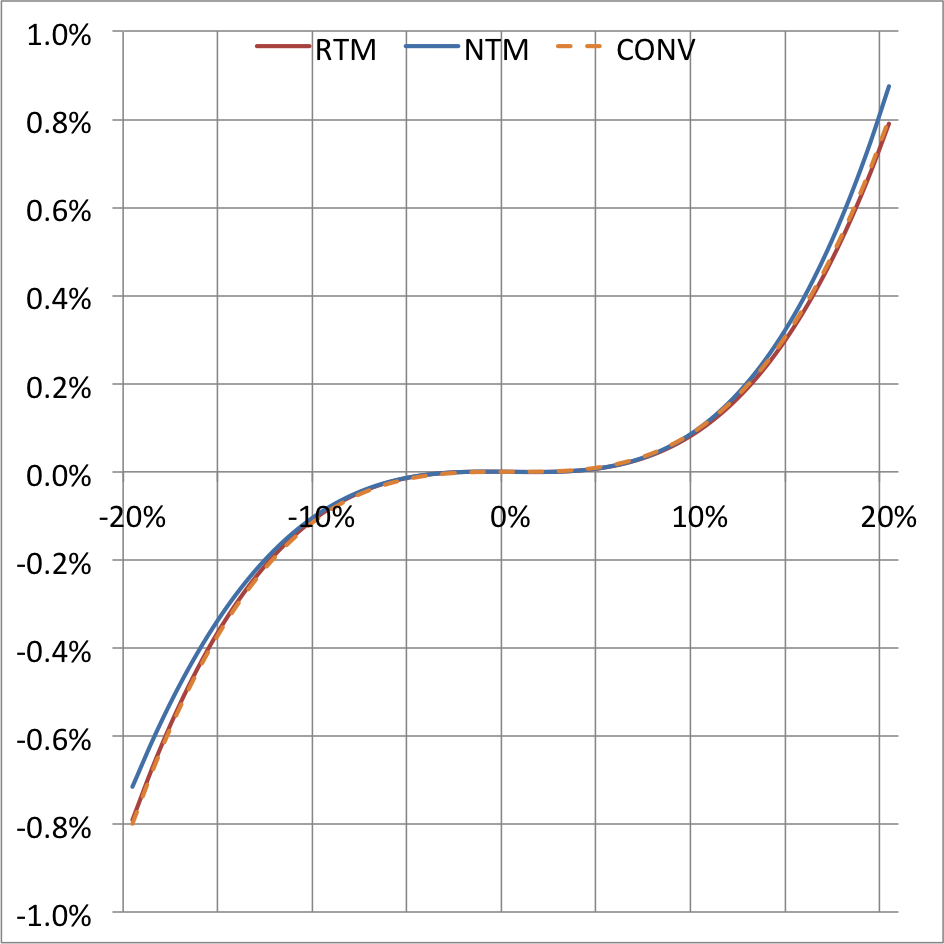}
		\includegraphics[width=0.49\linewidth,trim=0cm 0cm 0cm 0cm,clip=true]{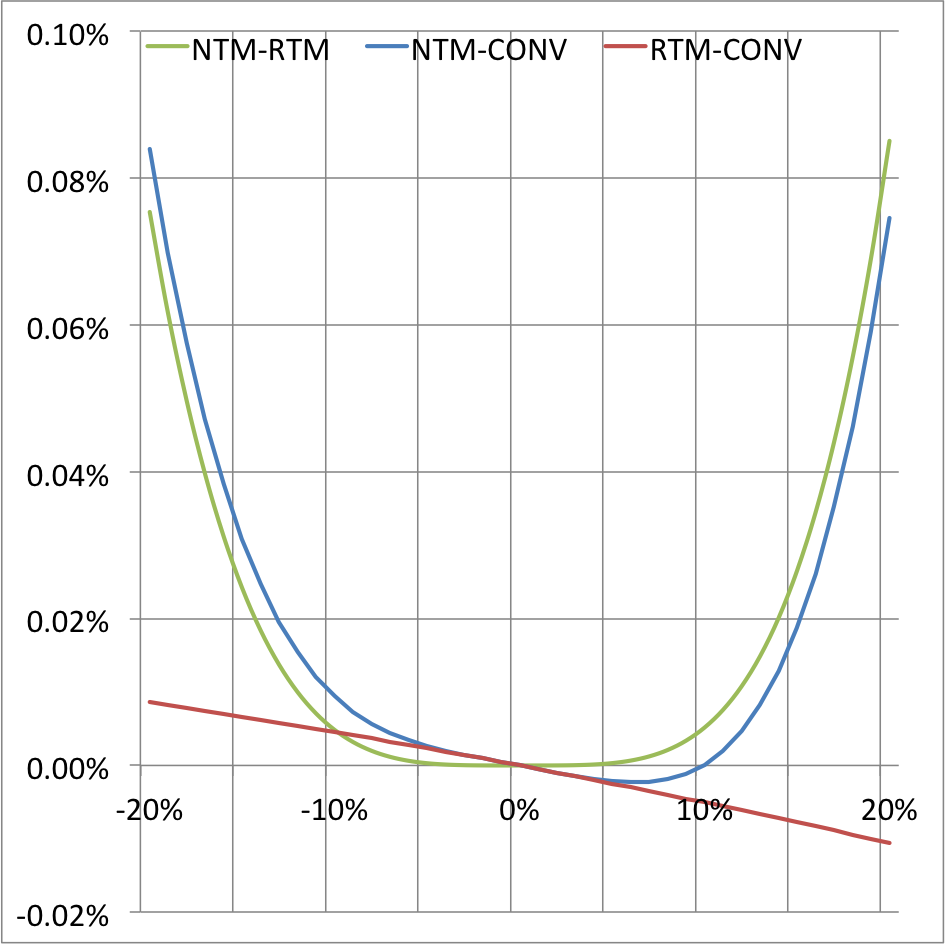}
		\par\end{centering}
	{\caption[Realised third-moment characteristics]{\small Comparison of \gls{rtm}, Neuberger's \gls{ntm} and the  \acrfull{clr}, i.e. $\left(y_s-y_r\right)^3$, for $s-t=1/250$ (daily monitoring) and $\sigma=20\%$ (implied volatility). Note that $v^\eta\approx v^{(2)}+v^{(3)}/3$, see Equation 14 in \cite{N12}. However, the graph for NTM does not change significantly for reasonable (or indeed unreasonable) values of the implied third moment so we set $v^{(3)}=0$ for simplicity. Again, the $x$-axis corresponds to $Y_s-Y_r=y_s-y_r-\tfrac{\sigma^2}{2}\left(s-r\right)$.}\label{fig:rtmcharacteristics}}
\end{figure}
The \gls{rtm} arises from two sources: the cubed change in price of the log contract (i.e. the short-term third moment) and the product of the change in the log contract and the change in conditional variance (i.e. leverage). Following the example of \cite{N12} (p.3430, last paragraph), if $F$ is a continuously-sampled continuous martingale, then the cubic term goes to zero and the only remaining source of long-term third moment is leverage. Both \gls{rtm} and \gls{ntm} require the replication of synthetic contracts and are therefore subject to a measurement error. The advantage of our definition is that, by \eqref{eq:cm}, the swap rate corresponds exactly to the third central moment of the log return distribution. 

In Figure \ref{fig:rtmcharacteristics} we compare the two realised third moment measures with the cubed  log return $\left(y_s-y_r\right)^3$, under the assumption that vanilla options with a continuum of strikes can be traded  and therefore that both the log and squared log contracts can be synthesised. As expected, the \gls{rtm} is closer to the \gls{clr} than \gls{ntm}. Note that \gls{rtm} is marginally below the cubed log return when $Y_s-Y_r$ is positive because the second, leverage term in RTM is  positive; and above it when $Y_s-Y_r$ is negative. The difference between \gls{rtm} and \gls{ntm} increases with the magnitude of $Y_s-Y_r$ and, for large positive or negative changes, \gls{ntm} is quite far above the conventional third moment.

Finally we set  $n=3$ in \eqref{eq:a} and \eqref{eq:b} to obtain a fourth-moment characteristic with  $a\left(\mathbf{u}\right)=3Y^4-6P^{(2)}Y^2+4P^{(3)}Y$ as well as $\mathbf{b}^\star\left(\mathbf{u}\right)=\left(-12Y^3+12P^{(2)}Y-4P^{(3)},6Y^2,-4Y\right)^\top$, and therefore (again, with $s-r=1/250$ for daily monitoring):
\begin{equation}\label{eq:rfm}\tag{RFM}
f\left(\mathbf{u}_r,\mathbf{u}_s\right)
=\left(Y_s-Y_r\right)^4+6v_s^{(2)}\left(Y_s-Y_r\right)^2+4\left(v_s^{(3)}-v_r^{(3)}\right)\left(Y_s-Y_r\right).
\end{equation}
Intuitively, this \gls{rfm} derives from three terms: the first is the fourth power of the change in price of the log contract (i.e. short-term fourth-moment); the second is a volatility clustering factor and the third is a leverage term. As with the third moment, the first term is zero if $F$ is a continuously-sampled continuous martingale. The volatility clustering factor is always positive and tends to decrease with residual time to maturity (for $s=T$ it is zero). If the implied distribution of the log price is symmetric then the leverage factor is zero, and then volatility clustering is the only source of kurtosis, if monitoring is continuous. The \gls{rfm} can be replicated by holding a fourth-power log contract as well as $6Y_r^2$ squared log contracts and shorting $12Y_r^3-12P_r^{(2)}Y_r+4P^{(3)}_r$ log contracts and $4Y_r$ cubed log contracts from time $r$ to time $s$. 


\begin{figure}[h!]
	\begin{centering}
		\includegraphics[width=0.49\linewidth,trim=0cm 0cm 0cm 0cm,clip=true]{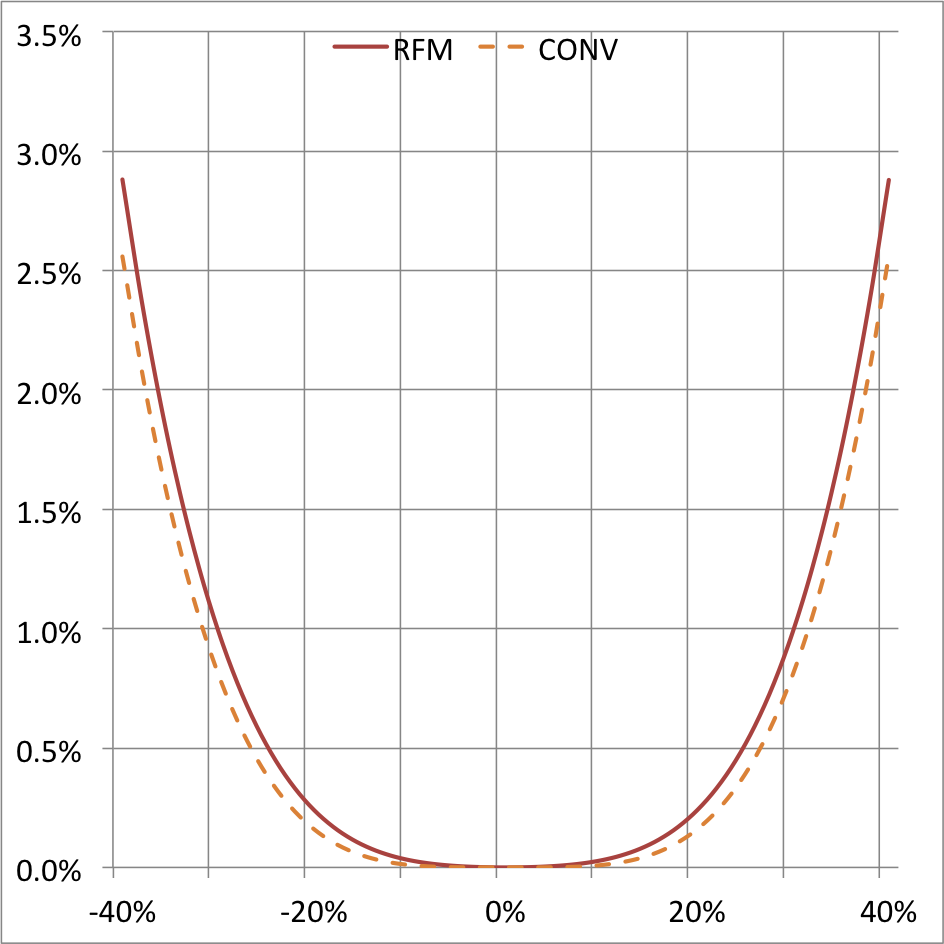}
		\includegraphics[width=0.49\linewidth,trim=0cm 0cm 0cm 0cm,clip=true]{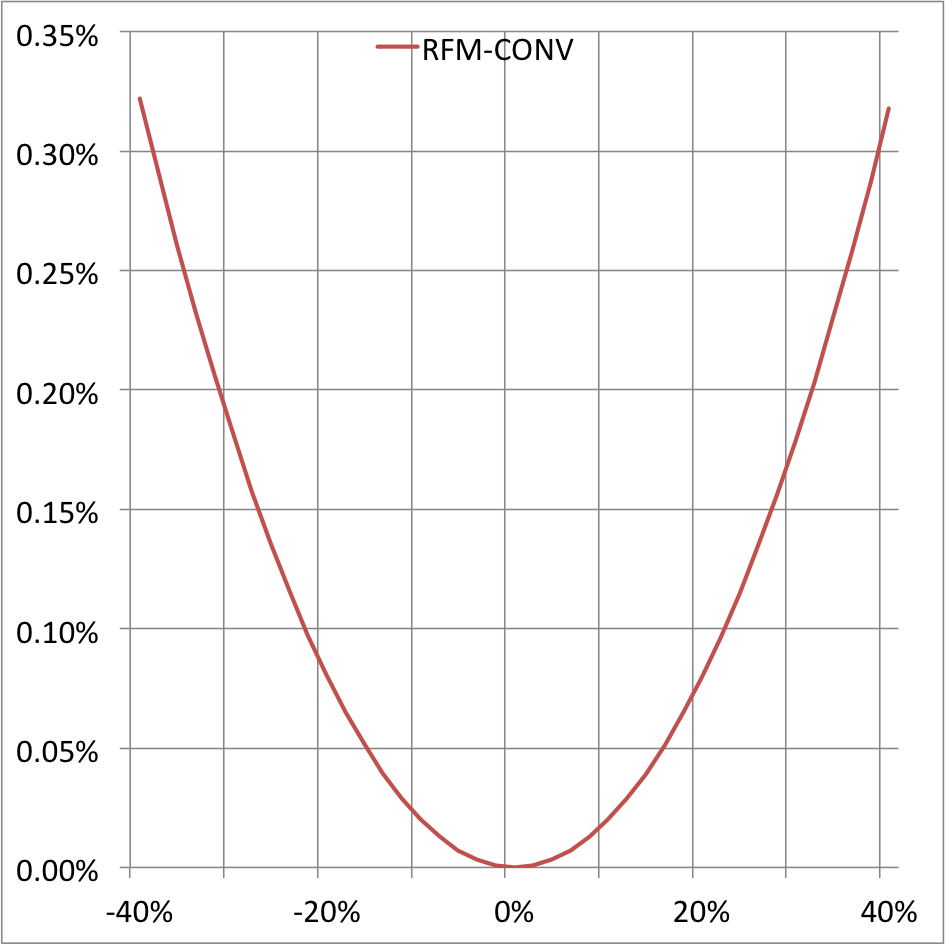}
		\par\end{centering}
	{\caption[Realised fourth-moment characteristic]{\small Comparison of \gls{rfm} with the \acrfull{qlr}, i.e. $\left(y_s-y_r\right)^4$, for $s-r=1/250$ (daily monitoring) and $\sigma=20\%$ (implied volatility). The graph for RFM does not change significantly for reasonable (or indeed unreasonable) values of implied third moment so we set $v^{(3)}=0$ for simplicity. Again, the $x$-axis corresponds to $Y_s-Y_r=y_s-y_r-\tfrac{\sigma^2}{2}\left(s-r\right)$.}\label{fig:rfmcharacteristics}}
\end{figure}
Figure \ref{fig:rfmcharacteristics} compares our realised fourth-moment characteristic with the conventional moment, i.e. the fourth power of the log return. By \eqref{eq:cm}, the former is constructed to: correspond exactly to an implied moment which captures the fourth central moment of log returns; to satisfy the \gls{ap} and thus be an unbiased estimator for the realised fourth moment, for any monitoring frequency; and to be the most efficient of these unbiased estimators under high-frequency monitoring. We argue that it has better properties than the conventional moment, which does not satisfy the \gls{ap} and, as a result, does not allow one to infer the values of long-term fourth moments from short-term observations. We note from the figure that there can be a substantial difference between our fourth moment definition and the conventional definition and that this difference increases with $|Y_s-Y_r|$.
   
\section{Conclusions}\label{sec:conc}
Our general  property encompasses two aggregation properties that were independently introduced by \cite{N12} and \cite{B14} as distinct examples, the former corresponding to a particular bivariate function and the latter being the univariate, martingale case of our general property. Our initial results are not confined to martingales, or even to log martingales, but in these cases we are able to define new, aggregating  characteristics which correspond exactly to higher moments of log returns.

To estimate a risk premium one takes the difference between the risk-neutral characteristic that is implied from traded options and the realised characteristic in the physical measure, which is typically derived from high-frequency historical data. But if the realised characteristic does not satisfy the aggregation property, the risk premium estimator will be biased -- just like the standard variance risk premium estimator. Moreover, the aggregation property allows one to derive unbiased estimators for the realised characteristic independent of the monitoring frequency. Unless the estimator satisfies the aggregation property, it is not possible to infer accurate values of long-term premia from short-term observations. 

While the unbiased property is independent of the monitoring frequency, one also needs to consider efficiency. Otherwise, the single historical time series that is observed on returns may, by chance, yield a  realised moment estimate which is far from its expectation, even though the estimator is unbiased. So within our vector space of unbiased estimators for higher moments we derive the most efficient, i.e. those with minimum conditional variance, where the efficiency of our selected estimators increases with the monitoring frequency.
  
We hope that this sets an agenda for further research. For instance,  when based upon our aggregating third and fourth moment characteristics, an empirical examination of the  determinants of high-moment risk premia may draw different conclusions to previous research. In particular, \cite{KNS13} extends the work of  \cite{CW09}, \cite{ELW10} and others on the determinants of the variance risk premium, only to conclude that the third-moment risk premium is very highly correlated with the variance risk premium. However, it may be that our characteristics which, unlike the aggregating third moment found by \cite{N12}, correspond exactly to the $n^{th}$ central moment of the implied distribution, do indeed yield diversified risk premia. Such a finding would be important for finance practitioners that seek new and profitable  forms of tradable risk.

\newpage
\singlespacing
\bibliographystyle{plainnat}
\bibliography{bibliography}

\newpage
\doublespacing
\section{Appendix}

\noindent {\bfseries Proof of Theorem 1:} Decomposing $f\left(\mathbf{u}_r,\mathbf{u}_s\right)$ using It\^o integrals yields
\begin{equation}\label{itoexpansion}
f\left(\mathbf{u}_r,\mathbf{u}_s\right)=\int_r^s\mathbf{J}\left(\mathbf{u}_r,\mathbf{u}_t\right)^{\top}d\mathbf{u}_t+\tfrac{1}{2}\int_r^s\text{tr}\left\{\mathbf{H}\left(\mathbf{u}_r,\mathbf{u}_t\right)d\langle\mathbf{u}\rangle_t\right\}.
\end{equation}
Applying \eqref{itoexpansion} to all terms in \eqref{eq:ap}, and writing $\mathbf{\hat{J}}_t=\mathbf{J}\left(\mathbf{u}_r,\mathbf{u}_t\right)-\mathbf{J}\left(\mathbf{u}_s,\mathbf{u}_t\right)$ as well as $\mathbf{\hat{H}}_t=\mathbf{H}\left(\mathbf{u}_r,\mathbf{u}_t\right)-\mathbf{H}\left(\mathbf{u}_s,\mathbf{u}_t\right)$,\footnote{For brevity of notation, we omit dependence of $\mathbf{\hat{J}}$ and $\mathbf{\hat{H}}$ on $r$ and $s$ because we shall see later that it is only the dependence on $t$ which is relevant for our proof.} yields
\begin{equation}\label{itoap}
\mathbbm{E}_r\left[\int_s^T\mathbf{\hat{J}}_t^{\top}d\mathbf{u}_t+\tfrac{1}{2}\int_s^T\text{tr}\left(\mathbf{\hat{H}}_t d\langle\mathbf{u}\rangle_t\right)\right]=0,
\end{equation}
Since $d\mathbf{x}_t=\boldsymbol\mu_tdt+\boldsymbol\Sigma_td\mathbf{w}_t$, It\^o's lemma applied to $\mathbf{u}_t=\mathbf{u}\left(t,\mathbf{x}_t\right)$ yields
\begin{equation}\label{udynamics}
d\mathbf{u}_t=\boldsymbol\lambda_tdt+ \boldsymbol\Delta_t\boldsymbol\Sigma_td\mathbf{w}_t,
\end{equation}
where 	
\begin{equation}\label{infinitesimalgenerator}
\boldsymbol\lambda_t=\boldsymbol\vartheta_t+\boldsymbol\Delta_t\boldsymbol\mu_t+\tfrac{1}{2}\left(\text{tr}\left\{\boldsymbol\Sigma_t^\top\boldsymbol\Gamma_{1t}\boldsymbol\Sigma_t\right\},\ldots,\text{tr}\left\{\boldsymbol\Sigma_t^\top\boldsymbol\Gamma_{nt}\boldsymbol\Sigma_t\right\}\right)^\top.
\end{equation}
The quadratic variation of $\mathbf{x}$ is $d\langle\mathbf{x}\rangle_t=\boldsymbol\Sigma_t\boldsymbol\Sigma_t^{\top}dt$, and so
\begin{equation}\label{uqv}
d\langle\mathbf{u}\rangle_t=\boldsymbol\Delta_t\boldsymbol\Sigma_t\boldsymbol\Sigma_t^\top\boldsymbol\Delta_t^{\top}dt.
\end{equation}
Inserting \eqref{udynamics}, \eqref{infinitesimalgenerator} and \eqref{uqv} in \eqref{itoap}, and noting that the stochastic integral w.r.t. $d\mathbf{w}$ vanishes under expectation, since $\mathbf{w}$ is a martingale, yields
\begin{equation}\label{condition}
\mathbbm{E}_r\left[\int_s^T\left\{\mathbf{\hat{J}}_t^{\top}\boldsymbol\lambda_t+\tfrac{1}{2}\text{tr}\left(\boldsymbol\Sigma_t^\top\boldsymbol\Delta_t^\top\mathbf{\hat{H}}_t\boldsymbol\Delta_t\boldsymbol\Sigma_t\right)\right\}dt\right]=0.
\end{equation}
Now we re-write \eqref{infinitesimalgenerator} into two terms, the first term $\boldsymbol\alpha_t$ containing parameters that are independent of $\boldsymbol\Sigma_t\boldsymbol\Sigma_t^\top$ and the second (trace) term containing the rest. For instance, $\boldsymbol\mu_t$ depends on $\boldsymbol\Sigma_t\boldsymbol\Sigma_t^\top$ when $\mathbf{x}_t$ is a log martingale, but it is zero when $\mathbf{x}_t$ is a martingale. The reason for this new decomposition of the drift term in \eqref{udynamics} is that we want to choose the parameters in a way that facilitates the derivation of necessary conditions for $(f;\mathbf{u})$ to satisfy \eqref{eq:ap}. Hence, write 
\begin{equation}\label{generatorspecification}
\boldsymbol\lambda_{t}=\boldsymbol\alpha_{t}+\tfrac{1}{2}\left(\text{tr}\left\{\boldsymbol\Sigma_t^\top\mathbfcal{B}_{1t}\boldsymbol\Sigma_t\right\},\ldots, \text{tr}\left\{\boldsymbol\Sigma_t^\top\mathbfcal{B}_{nt}\boldsymbol\Sigma_t\right\}\right)^\top,
\end{equation}
where $\boldsymbol\alpha_t=\boldsymbol\alpha\left(t,\mathbf{x}_t\right)\in\mathbbm{R}^n$ and we can assume w.l.o.g. that $\mathbfcal{B}_{it}=\mathbfcal{B}_i\left(t,\mathbf{x}_t\right)\in\mathbbm{R}^{n\times n}$ are symmetric matrices, for $i\in\{1,\ldots,n\}$.\footnote{This is because they only appear within the trace operator, and if $\mathbfcal{B}_{it}$ were not symmetric we can always find another, symmetric matrix with the same trace as  $\boldsymbol\Sigma_t^\top\mathbfcal{B}_{it}\boldsymbol\Sigma_t$.}

Next we show that non-trivial solutions arise iff $\boldsymbol{\alpha}= \mathbf{0}$. That is, the only non-zero drift $\boldsymbol{\lambda}_t$ that yields non-trivial solutions must depend on  $\boldsymbol\Sigma_t\boldsymbol\Sigma_t^{\top}$, as would e.g. be the case when $\mathbf{u}$ is a log martingale. To see this, insert \eqref{generatorspecification} in \eqref{condition}, to obtain 
\begin{equation}\label{conditionlambda}
\mathbbm{E}_r\left[\int_s^T\left\{\mathbf{\hat{J}}_t^{\top}\boldsymbol\alpha_t+\tfrac{1}{2}\text{tr}\left(\boldsymbol\Sigma_t^\top\left\{\hat{J}_{1t}\mathbfcal{B}_{1t}+\ldots+\hat{J}_{nt}\mathbfcal{B}_{nt}+\boldsymbol\Delta_t^\top\mathbf{\hat{H}}_t\boldsymbol\Delta_t\right\}\boldsymbol\Sigma_t\right)\right\}dt\right]=0,
\end{equation}
where $\mathbf{\hat{J}}=\left(\hat{J}_1,\ldots,\hat{J}_n\right)^\top$. Now consider the spectral decomposition
\begin{equation}\label{spectraldecomposition}
\hat{J}_{1t}\mathbfcal{B}_{1t}+\ldots+\hat{J}_{nt}\mathbfcal{B}_{nt}+\boldsymbol\Delta_t^\top\mathbf{\hat{H}}_t\boldsymbol\Delta_t=\mathbf{W}_t\mathbf{V}_t\mathbf{W}_t^\top,
\end{equation}
where $\mathbf{V}_t\in\mathbbm{R}^{n\times n}$ is a diagonal matrix of eigenvalues and the columns of $\mathbf{W}_t\in\mathbbm{R}^{n\times n}$ contain the corresponding eigenvectors, which are orthogonal by definition and hence $\mathbf{W}_t^\top\mathbf{W}_t=\mathbf{I}$ (matrix identity). Now we set a specific drift and a specific volatility in our diffusion. For the volatility, we suppose that
\begin{equation}\label{volatilityspecification}
\boldsymbol\Sigma_t=\sigma\exp\left(\tfrac{\xi}{2}\mathbf{W}_t\mathbf{V}_t\mathbf{W}_t^\top\right)=\sigma\mathbf{W}_t\exp\left(\tfrac{\xi}{2}\mathbf{V}_t\right)\mathbf{W}_t^\top,
\end{equation}
for some real  $\xi$ and $\sigma \ge 0$. For the drift we assume
\begin{equation}\label{driftspecification}
\boldsymbol\alpha_t=\alpha\mathbf{\hat{J}}_t,
\end{equation}
for some $\alpha\in\mathbbm{R}$. Inserting \eqref{spectraldecomposition}, \eqref{volatilityspecification} and \eqref{driftspecification} into \eqref{conditionlambda}, differentiating w.r.t. $T$ and $\xi$ using Leibnitz' rule for differentiation, and applying the cyclic property of the trace operator yields
$$\alpha\mathbbm{E}_r\left[\mathbf{\hat{J}}_t^\top\mathbf{\hat{J}}_t\right]=0, \mbox{ and }\,
\sigma^2\mathbbm{E}_r\left[\text{tr}\left(\mathbf{V}_t^2\exp\left\{\xi\mathbf{V}_t\right\}\right)\right]=0.$$ Note that each term in $\mathbf{\hat{J}}_t^\top\mathbf{\hat{J}}_t$ is $\ge 0$.  Hence, $\alpha\mathbbm{E}_r\left[\mathbf{\hat{J}}_t^\top\mathbf{\hat{J}}_t\right]=0 \Rightarrow \alpha=0$ or $\mathbf{\hat{J}}_t=\mathbf{0}$, or both; Similarly, each term in $\text{tr}\left(\mathbf{V}_t^2\exp\left\{\xi\mathbf{V}_t\right\}\right)$ is $\ge 0$. Hence, either $\sigma=0$ or $\mathbf{V}_t=\mathbf{0}$, or both. 
Setting $\sigma=0$ corresponds to a deterministic process, and $\mathbf{\hat{J}}_t=\mathbf{0}$ yields the trivial solution $f\left(\mathbf{u}_r,\mathbf{u}_s\right)=a\left(\mathbf{u}_s\right)-a\left(\mathbf{u}_r\right)$, since $f\left(\mathbf{u},\mathbf{u}\right)=0$ for all $\mathbf{u}\in\mathbbm{R}^n$. So we assume that $\alpha=0$ and $\sigma>0$, which implies that $\boldsymbol\alpha_t=\mathbf{0}$ and $\mathbf{V}_t=\mathbf{0}$, for all $t$.

First we show that the condition $\mathbf{V}_t=\mathbf{0}$ implies \eqref{eq:pde1}. To see this, insert $\mathbf{V}_t=\mathbf{0}$ in \eqref{spectraldecomposition} to yield:
\begin{equation}\label{eq:pdep}
\hat{J}_{1t}\mathbfcal{B}_{1t}+\ldots+\hat{J}_{nt}\mathbfcal{B}_{nt}+\boldsymbol\Delta_t^\top\mathbf{\hat{H}}_t\boldsymbol\Delta_t=\mathbf{0}.
\end{equation} 
Rearranging \eqref{eq:pdep}, and setting $\mathbf{C}_{it}=-\left(\boldsymbol\Delta_t^\top\right)^{-1}\mathbfcal{B}_{it}\boldsymbol\Delta_t^{-1}$, which is symmetric by definition, yields
\begin{equation*}\label{eq:pdep2}
\mathbf{\hat{H}}_t=-\hat{J}_{1t}\mathbf{C}_{1t}-\ldots-\hat{J}_{nt}\mathbf{C}_{nt}.
\end{equation*}
Expanding $\mathbf{\hat{H}}$ and $\mathbf{\hat{J}}$ yields
\begin{equation*}
\mathbf{H}\left(\mathbf{u}_r,\mathbf{u}_t\right)-\sum_{i=1}^nJ_i\left(\mathbf{u}_r,\mathbf{u}_t\right)\mathbf{C}_{it}=\mathbf{H}\left(\mathbf{u}_s,\mathbf{u}_t\right)-\sum_{i=1}^nJ_i\left(\mathbf{u}_s,\mathbf{u}_t\right)\mathbf{C}_{it}.
\end{equation*}
In other words, the expression on the right (and the left) depends only on $\mathbf{u}_t$, not on $\mathbf{u}_r$ or $\mathbf{u}_s$. Therefore there must be some function $a\in\mathcal{C}^2$ which satisfies
\begin{equation*}
	\mathbf{H}\left(\mathbf{u}_r,\mathbf{u}_t\right)-\sum_{i=1}^nJ_i\left(\mathbf{u}_r,\mathbf{u}_t\right)\mathbf{C}_{it}=\mathbf{H}^a\left(\mathbf{u}_t\right)-\sum_{i=1}^n J_i^a\left(\mathbf{u}_t\right)\mathbf{C}_{it}.
\end{equation*}
Then \eqref{eq:pde1} follows on setting $t=s$:
\begin{equation*}
\mathbf{H}\left(\mathbf{u}_r,\mathbf{u}_s\right)-\mathbf{H}^a\left(\mathbf{u}_s\right)=\sum_{i=1}^n\left\{J_i\left(\mathbf{u}_r,\mathbf{u}_s\right)-J_i^a\left(\mathbf{u}_s\right)\right\}\mathbf{C}_{is}.
\end{equation*}
Note that $\mathbf{x}_t$ can be expressed as a function of $\mathbf{u}_t$ if $\boldsymbol\Delta_t$ is invertible, and the l.h.s. of the above equation depends on $t$ only through $\mathbf{u}_t$, hence $\mathbf{C}_{it}=\mathbf{C}_i\left(\mathbf{u}_t\right)$. Finally, setting $\boldsymbol{\alpha}_t=\mathbf{0}$ in \eqref{generatorspecification}, substituting $\mathbfcal{B}_{it}=-\boldsymbol\Delta_t^\top\mathbf{C}_{it}\boldsymbol\Delta_t$, and equating this with \eqref{infinitesimalgenerator}, yields \eqref{eq:pde2}.\qed\\

\noindent {\bfseries Proof of Corollary 1:} Define
\begin{equation*}
\mathbf{y}_s=\mathbf{u}\left(s,\mathbf{x}_s\right)+\tfrac{1}{2}\int_t^s\left(\text{tr}\left\{\boldsymbol\Sigma_\tau^\top\boldsymbol\Delta_\tau^\top\mathbf{C}_{1\tau}\boldsymbol\Delta_\tau\boldsymbol\Sigma_\tau\right\},\ldots,\text{tr}\left\{\boldsymbol\Sigma_\tau^\top\boldsymbol\Delta_\tau^\top\mathbf{C}_{n\tau}\boldsymbol\Delta_\tau\boldsymbol\Sigma_\tau\right\}\right)^\top d\tau,
\end{equation*}
so that $\mathbf{y}_t=\mathbf{u}\left(t,\mathbf{x}_t\right)$. By It\^o's lemma, 
$$
d\mathbf{y}_s=d\mathbf{u}_s+\tfrac{1}{2}\left(\text{tr}\left\{\boldsymbol\Sigma_s^\top\boldsymbol\Delta_s^\top\mathbf{C}_{1s}\boldsymbol\Delta_s\boldsymbol\Sigma_s\right\},\ldots,\text{tr}\left\{\boldsymbol\Sigma_s^\top\boldsymbol\Delta_s^\top\mathbf{C}_{ns}\boldsymbol\Delta_s\boldsymbol\Sigma_s\right\}\right)^\top ds.
$$
Substituting for $d\mathbf{u}_s$ using \eqref{udynamics}, and \eqref{generatorspecification} with $\boldsymbol{\alpha}_s=\mathbf{0}, \, \forall\, s$, yields $d\mathbf{y}_s=\boldsymbol\Delta_s\boldsymbol\Sigma_sd\mathbf{w}_s,$
i.e. $\mathbf{y}_s$ is a martingale. Hence
\begin{eqnarray*}
\mathbf{y}_t&=&\mathbbm{E}_t\left[\mathbf{y}_T\right]=\mathbbm{E}_t\left[\boldsymbol\psi\left(\mathbf{x}_T\right)+\tfrac{1}{2}\int_t^T\left(\text{tr}\left\{\boldsymbol\Sigma_\tau^\top\boldsymbol\Delta_\tau^\top\mathbf{C}_{1\tau}\boldsymbol\Delta_\tau\boldsymbol\Sigma_\tau\right\},\ldots,\text{tr}\left\{\boldsymbol\Sigma_\tau^\top\boldsymbol\Delta_\tau^\top\mathbf{C}_{n\tau}\boldsymbol\Delta_\tau\boldsymbol\Sigma_\tau\right\}\right)^\top d\tau\right]\\
&=&\mathbf{u}\left(t,\mathbf{x}_t\right),
\end{eqnarray*}
where we have substituted the boundary condition $\mathbf{u}\left(T,\mathbf{x}_T\right)=\boldsymbol\psi\left(\mathbf{x}_T\right)$.\qed\\

\noindent The proof of Theorem 2 requires the following:\\

\noindent {\bfseries Lemma:} Let $\mathbf{C}_i=\mathbf{Q}\mathbf{D}_{i}\mathbf{Q}^\top$ and $\sum_{i=1}^n\mathbf{C}_i=\left(\mathbf{C}_1\mathbf{1},\ldots,\mathbf{C}_n\mathbf{1}\right)$. Then also
\begin{equation*}
\sum_{i=1}^n\mathbf{C}_ik_i=\left(\mathbf{C}_1\mathbf{k},\ldots,\mathbf{C}_n\mathbf{k}\right)=\left(\mathbf{C}_1^\top\mathbf{k},\ldots,\mathbf{C}_n^\top\mathbf{k}\right)^\top,
\end{equation*}
for all $\mathbf{k}=(k_1,\ldots,k_n)^\top\in\mathbf{R}^n$. That is, the third order tensor $\left(\mathbf{C}_1,\ldots,\mathbf{C}_n\right)$ is symmetric.\\

\noindent {\bfseries Proof of Lemma:} Let $\mathbf{Q}=\left(\mathbf{q}_1,\ldots,\mathbf{q}_n\right)^\top$ and $\mathbf{D}_i=\text{diag}\left(\mathbf{D}\mathbf{q}_i\right)$. Clearly $\mathbf{D}$ exists, and it is unique if $\mathbf{C}_i$ are linearly independent. Combining this with $\sum_{i=1}^n\mathbf{C}_i=\left(\mathbf{C}_1\mathbf{1},\ldots,\mathbf{C}_n\mathbf{1}\right)$ yields
\begin{equation*}
\sum_{i=1}^n\mathbf{Q}\, \text{diag}\left(\mathbf{D}\mathbf{q}_i\right)\mathbf{Q}^\top=\left(\mathbf{Q}\, \text{diag}\left(\mathbf{D}\mathbf{q}_1\right)\mathbf{Q}^\top\mathbf{1},\ldots,\mathbf{Q}\, \text{diag}\left(\mathbf{D}\mathbf{q}_n\right)\mathbf{Q}^\top\mathbf{1}\right).
\end{equation*}
Since $\, \text{diag}\left(\mathbf{D}\mathbf{q}_i\right)\mathbf{Q}^\top\mathbf{1}=\, \text{diag}\left(\mathbf{Q}^\top\mathbf{1}\right)\mathbf{D}\mathbf{q}_i$ we have
\begin{equation*}
\mathbf{D}=\sum_{i=1}^n\, \text{diag}\left(\mathbf{D}\mathbf{q}_i\right)\, \text{diag}\left(\mathbf{Q}^\top\mathbf{1}\right)^{-1}=\, \text{diag}\left(\mathbf{D}\mathbf{1}\right).
\end{equation*}
Therefore $\mathbf{D}$ must be diagonal.  Then
\begin{eqnarray*}
	\sum_{i=1}^n\mathbf{C}_ik_i&=&\sum_{i=1}^n\mathbf{Q}\, \text{diag}\left(\mathbf{D}\mathbf{q}_i\right)\mathbf{Q}^\top_ik_i=\mathbf{Q}\, \text{diag}\left(\mathbf{D}\mathbf{Q}^\top\mathbf{k}\right)\left(\, \text{diag}\left\{\mathbf{q}_1\right\}\mathbf{1},\ldots,\, \text{diag}\left\{\mathbf{q}_n\right\}\mathbf{1}\right)\\
	&=&\left(\mathbf{Q}\, \text{diag}\left(\mathbf{D}\mathbf{q}_1\right)\mathbf{Q}^\top\mathbf{k},\ldots,\mathbf{Q}\, \text{diag}\left(\mathbf{D}\mathbf{q}_n\right)\mathbf{Q}^\top\mathbf{k}\right)=\left(\mathbf{C}_1\mathbf{k},\ldots,\mathbf{C}_n\mathbf{k}\right),
\end{eqnarray*}
where we have used that $\text{diag}\left(\mathbf{D}\mathbf{Q}^\top\mathbf{k}\right)\, \text{diag}\left(\mathbf{q}_i\right)=\, \text{diag}\left(\mathbf{D}\mathbf{q}_i\right)\, \text{diag}\left\{\mathbf{Q}^\top\mathbf{k}\right\}$. Finally, since $\mathbf{C}_i$ are symmetric, we have $\left(\mathbf{C}_1\mathbf{k},\ldots,\mathbf{C}_n\mathbf{k}\right)=\left(\mathbf{C}_1^\top\mathbf{k},\ldots,\mathbf{C}_n^\top\mathbf{k}\right)^\top$.
\qed\\

\noindent {\bfseries Proof of Theorem 2:} First note that symmetric $\mathbf{C}_{it}$ commute for all $i$ and $t$ iff $\mathbf{C}_{it}=\mathbf{Q}\mathbf{D}_{it}\mathbf{Q}^\top$. See \cite{HJ85}, p.52. Then we have first and second partial derivatives of $\mathbf{m}$ w.r.t. components of $\mathbf{u}$:
\begin{equation*}
\mathbf{J}^m\left(\mathbf{u}_t\right)=\exp\left\{\sum_{i=1}^n\int_0^{u_{it}}\mathbf{C}_i\left(\mathbf{u}\right)du_i\right\}=\prod_{i=1}^n\exp\left\{\int_0^{u_{it}}\mathbf{C}_i\left(\mathbf{u}\right)du_i\right\},
\end{equation*}
and
$\mathbf{H}_i^m\left(\mathbf{u}_t\right)=\mathbf{C}_{it}\mathbf{J}^m\left(\mathbf{u}_t\right)=\mathbf{J}^m\left(\mathbf{u}_t\right)\mathbf{C}_{it}$. 
Now consider
\begin{eqnarray*}
	&&\left(\tfrac{\partial}{\partial u_{1s}},\ldots,\tfrac{\partial}{\partial u_{ns}}\right)^\top\left(\left\{\mathbf{J}\left(\mathbf{u}_r,\mathbf{u}_s\right)-\mathbf{J}^a\left(\mathbf{u}_s\right)\right\}^\top\mathbf{J}^m\left(\mathbf{u}_s\right)^{-1}\right)\\
	&=&\left(\mathbf{H}\left(\mathbf{u}_r,\mathbf{u}_s\right)-\mathbf{H}^a\left(\mathbf{u}_s\right)-\left\{\mathbf{J}\left(\mathbf{u}_r,\mathbf{u}_s\right)-\mathbf{J}^a\left(\mathbf{u}_s\right)\right\}^\top\left(\mathbf{C}_{1s},\ldots,\mathbf{C}_{ns}\right)\right)\mathbf{J}^m\left(\mathbf{u}_s\right)^{-1}\\
	&=&\left(\mathbf{H}\left(\mathbf{u}_r,\mathbf{u}_s\right)-\mathbf{H}^a\left(\mathbf{u}_s\right)-\sum_{i=1}^n\left\{J_i\left(\mathbf{u}_r,\mathbf{u}_s\right)-J_i^a\left(\mathbf{u}_s\right)\right\}\mathbf{C}_{is}\right)\mathbf{J}^m\left(\mathbf{u}_s\right)^{-1}=\mathbf{0},
\end{eqnarray*}
where we apply the Lemma in the second line and \eqref{eq:pde1} in the third line, and hence
\begin{equation*}
\left\{\mathbf{J}\left(\mathbf{u}_r,\mathbf{u}_s\right)-\mathbf{J}^a\left(\mathbf{u}_s\right)\right\}^\top\mathbf{J}^m\left(\mathbf{u}_s\right)^{-1}=\mathbf{b}\left(\mathbf{u}_r\right)^\top,
\end{equation*}
for some differentiable $\mathbf{b}:\mathbbm{R}^n\rightarrow\mathbbm{R}^n$ (note that $\mathbf{b}\left(\mathbf{u}\right)=\mathbf{J}\left(\mathbf{u},\mathbf{u}\right)-\mathbf{J}^a\left(\mathbf{u}\right)$). Integration yields
\begin{equation*}
	f\left(\mathbf{u}_r,\mathbf{u}_s\right)=a\left(\mathbf{u}_s\right)-a\left(\mathbf{u}_r\right)+\mathbf{b}\left(\mathbf{u}_r\right)^\top\left\{\mathbf{m}\left(\mathbf{u}_s\right)-\mathbf{m}\left(\mathbf{u}_r\right)\right\},
\end{equation*}
where we have used that $f\left(\mathbf{u},\mathbf{u}\right)=0$. We can assume w.l.o.g. that $a\left(\mathbf{0}\right)=0$. Furthermore It\^o's formula applied to $\mathbf{m}$ yields
\begin{equation*}
d\mathbf{m}\left(\mathbf{u}_t\right)=\mathbf{J}^m\left(\mathbf{u}_t\right)d\mathbf{u}_t+\tfrac{1}{2}\mathbf{J}^m\left(\mathbf{u}_t\right)\left(\text{tr}\left\{\mathbf{C}_{1t}d\left\langle\mathbf{u}\right\rangle_t\right\},\ldots,\text{tr}\left\{\mathbf{C}_{nt}d\left\langle\mathbf{u}\right\rangle_t\right\}\right)^\top=\mathbf{J}^m\left(\mathbf{u}_t\right)\boldsymbol\Delta_t\boldsymbol\Sigma_td\mathbf{w}_t,
\end{equation*}
and therefore $\mathbf{m}$ is a multivariate martingale.\qed\\

\noindent {\bfseries Proof of Corollary 2:} Note that
\begin{equation}\label{mlogmartingale}
\mathbf{m}\left(\mathbf{u}_t\right)=\int_0^{\mathbf{u}_t}\exp\left\{\sum_{i=1}^n\mathbf{C}_iu_i\right\}d\mathbf{u}=\left(\sum_{i=1}^n\mathbf{C}_ik_i\right)^{-1}\left(\exp\left\{\sum_{i=1}^n\mathbf{C}_iu_{it}\right\}-\mathbf{I}\right)\mathbf{k},
\end{equation}
for some $\mathbf{k}$ s.t. $\sum_{i=1}^n\mathbf{C}_ik_i$ is invertible. First we assume w.l.o.g. that $\mathbf{k}=\mathbf{1}$ and therefore
\begin{equation*}
\mathbf{m}\left(\mathbf{u}_T\right)=\left(\sum_{i=1}^n\mathbf{C}_i\right)^{-1}\left(\exp\left\{\sum_{i=1}^n\mathbf{C}_i\psi_i\left(\mathbf{x}_T\right)\right\}-\mathbf{I}\right)\mathbf{1}.
\end{equation*}
Now $\mathbf{m}$ is a martingale and we have $\mathbf{m}\left(\mathbf{u}_t\right)=\mathbbm{E}_t\left[\mathbf{m}\left(\mathbf{u}_T\right)\right]$, i.e.
\begin{equation}\label{mlogmartingaleboundary}
\mathbf{m}\left(\mathbf{u}_t\right)=\left(\sum_{i=1}^n\mathbf{C}_ik_i\right)^{-1}\mathbbm{E}_t\left[\exp\left\{\sum_{i=1}^n\mathbf{C}_i\psi_i\left(\mathbf{x}_T\right)\right\}-\mathbf{I}\right]\mathbf{1}.
\end{equation}
Secondly, since $\sum_{i=1}^n\mathbf{C}_ik_i\mathbf{m}=\sum_{i=1}^n\mathbf{C}_im_i\mathbf{k}$ by the Lemma, we can write $\mathbf{u}$ as a function of $\mathbf{m}$ rather than the converse, using the equivalent expression to \eqref{mlogmartingale}, i.e.
\begin{equation}\label{mdual}
\mathbf{u}_t=\left(\sum_{i=1}^n\mathbf{C}_i\right)^{-1}\ln\left(\mathbf{I}+\sum_{i=1}^n\mathbf{C}_im_{it}\right)\mathbf{1},
\end{equation}
where the choice of $\mathbf{1}$ is again arbitrary. Finally, inserting \eqref{mlogmartingaleboundary} in \eqref{mdual} and once again making use of the Lemma yields
\begin{equation*}
\mathbf{u}\left(t,\mathbf{x}_t\right)=\left(\sum_{i=1}^n\mathbf{C}_i\right)^{-1}\ln\left(\mathbbm{E}_t\left[\exp\left\{\sum_{i=1}^n\mathbf{C}_i\psi_i\left(\mathbf{x}_T\right)\right\}\right]\right)\mathbf{1},
\end{equation*}
the log of a martingale satisfying the boundary condition $\mathbf{u}_T=\boldsymbol\psi\left(\mathbf{x}_T\right)$.\qed\\

\noindent {\bfseries Proof of Corollary 3:} We can either derive \eqref{eq:msoln} as the limit for $\mathbf{C}_i\rightarrow\mathbf{0}$ in Theorem 2 or, alternatively, directly from \eqref{eq:pde1} in Theorem 1. For the former derivation, consider a first-order Taylor expansion of the solution
\begin{equation*}
	f\left(\mathbf{u}_r,\mathbf{u}_s\right)=a\left(\mathbf{u}_s\right)-a\left(\mathbf{u}_r\right)+\mathbf{b}\left(\mathbf{u}_r\right)^\top\left(\sum_{i=1}^n\mathbf{C}_i\right)^{-1}\sum_{i=1}^n\mathbf{C}_i\left(u_{is}-u_{ir}\right)\mathbf{1}+O\left(\left|\sum_{i=1}^n\mathbf{C}_i\right|\right),
\end{equation*}
where $\sum_{i=1}^n\mathbf{C}_i$ cancels out in the $\mathbf{b}$ term (after applying the Lemma) and the higher orders vanish as $\mathbf{C}_i\rightarrow\mathbf{0}$, $i=1,\ldots,n$. For the latter derivation, note that setting all $\mathbf{C}_{is}=\mathbf{0}$ in Theorem 1 yields $\mathbf{H}\left(\mathbf{u}_r,\mathbf{u}_s\right)=\mathbf{H}^a\left(\mathbf{u}_s\right)$. Then
\begin{equation*}
\int_{\mathbf{u}_r}^{\mathbf{u}_s}\left[\int_{\mathbf{u}_r}^{\mathbf{u}_t}d\mathbf{u}_s^\top\mathbf{H}\left(\mathbf{u}_r,\mathbf{u}_s\right)\right]d\mathbf{u}_t=\int_{\mathbf{u}_r}^{\mathbf{u}_s}\left[\int_{\mathbf{u}_r}^{\mathbf{u}_t}d\mathbf{u}_s^\top\mathbf{H}^a\left(\mathbf{u}_s\right)\right]d\mathbf{u}_t,
\end{equation*}
and integrating yields
\begin{equation*}
\int_{\mathbf{u}_r}^{\mathbf{u}_s}\left[\mathbf{J}\left(\mathbf{u}_r,\mathbf{u}_t\right)-\mathbf{J}\left(\mathbf{u}_r,\mathbf{u}_r\right)\right]^\top d\mathbf{u}_t=\int_{\mathbf{u}_r}^{\mathbf{u}_s}\left[\mathbf{J}^a\left(\mathbf{u}_t\right)-\mathbf{J}^a\left(\mathbf{u}_r\right)\right]^\top d\mathbf{u}_t.
\end{equation*}
Integrating the above once again and using $f\left(\mathbf{u},\mathbf{u}\right)=0$  yields the solution
\begin{equation*}
f\left(\mathbf{u}_r,\mathbf{u}_s\right)=a\left(\mathbf{u}_s\right)-a\left(\mathbf{u}_r\right)+\mathbf{b}\left(\mathbf{u}_r\right)^\top\left(\mathbf{u}_s-\mathbf{u}_r\right),
\end{equation*}
where $\mathbf{b}\left(\mathbf{u}\right)=\mathbf{J}\left(\mathbf{u},\mathbf{u}\right)-\mathbf{J}^a\left(\mathbf{u}\right)$, so $\mathbf{b}:\mathbbm{R}^n\rightarrow\mathbbm{R}^n$ must be differentiable. Finally, setting $\mathbf{C}_{it}=\mathbf{0}$ for $i=1,\ldots,n$ in Corollary 1 yields $\mathbf{u}_t=\mathbbm{E}_t\left[\boldsymbol\psi\left(\mathbf{x}_T\right)\right]$.\qed\\

\noindent {\bfseries Proof of Theorem 3:} For $t\in\left\{t_0,\ldots,t_N\right\}$ we are interested in minimising the expression
\begin{eqnarray*}
&&\mathbbm{E}_t\left[\left\{\sum_{i=1}^Nf\left(\mathbf{u}_{i-1},\mathbf{u}_i\right)-\mathbbm{E}_t\left[\sum_{i=1}^Nf\left(\mathbf{u}_{i-1},\mathbf{u}_i\right)\right]\right\}^2\right]\\
&=&\mathbbm{E}_t\left[\left\{a\left(\mathbf{u}_T\right)-\mathbbm{E}_t\left[a\left(\mathbf{u}_T\right)\right]+\sum_{t_i>t}\mathbf{b}\left(\mathbf{u}_{i-1}\right)^\top\left(\mathbf{u}_i-\mathbf{u}_{i-1}\right)\right\}^2\right]\\
&=&\mathbbm{E}_t\left[a\left(\mathbf{u}_T\right)^2\right]-\mathbbm{E}_t\left[a\left(\mathbf{u}_T\right)\right]^2+2\sum_{t_i>t}\mathbbm{E}_t\left[a\left(\mathbf{u}_T\right)\mathbf{b}\left(\mathbf{u}_{i-1}\right)^\top\left(\mathbf{u}_i-\mathbf{u}_{i-1}\right)\right]\\
&&+\sum_{t_i>t}\mathbbm{E}_t\left[\left\{\mathbf{b}\left(\mathbf{u}_{i-1}\right)^\top\left(\mathbf{u}_i-\mathbf{u}_{i-1}\right)\right\}^2\right]\\
&=&\mathbbm{E}_t\left[a\left(\mathbf{u}_T\right)^2\right]-\mathbbm{E}_t\left[a\left(\mathbf{u}_T\right)\right]^2+2\sum_{t_i>t}\mathbbm{E}_t\left[\mathbf{b}\left(\mathbf{u}_{i-1}\right)^\top\mathbbm{E}_{t_{i-1}}\left[a\left(\mathbf{u}_T\right)\left(\mathbf{u}_i-\mathbf{u}_{i-1}\right)\right]\right]\\
&&+\sum_{t_i>t}\mathbbm{E}_t\left[\mathbf{b}\left(\mathbf{u}_{i-1}\right)^\top\mathbbm{E}_{t_{i-1}}\left[\left(\mathbf{u}_i-\mathbf{u}_{i-1}\right)\left(\mathbf{u}_i-\mathbf{u}_{i-1}\right)^\top\right]\mathbf{b}\left(\mathbf{u}_{i-1}\right)\right]\\
&=&\mathbbm{E}_t\left[a\left(\mathbf{u}_T\right)^2\right]-\mathbbm{E}_t\left[a\left(\mathbf{u}_T\right)\right]^2+\sum_{t_i>t}\mathbbm{E}_t\left[2\mathbf{b}\left(\mathbf{u}_{i-1}\right)^\top\boldsymbol\omega_{i-1}+\mathbf{b}\left(\mathbf{u}_{i-1}\right)^\top\boldsymbol\Omega_{i-1}\mathbf{b}\left(\mathbf{u}_{i-1}\right)\right],
\end{eqnarray*}
with $\boldsymbol\omega_{i-1}=\mathbbm{E}_{t_{i-1}}\left[a\left(\mathbf{u}_T\right)\left(\mathbf{u}_i-\mathbf{u}_{i-1}\right)\right]$ and $\boldsymbol\Omega_{i-1}=\mathbbm{E}_{t_{i-1}}\left[\left(\mathbf{u}_i-\mathbf{u}_{i-1}\right)\left(\mathbf{u}_i-\mathbf{u}_{i-1}\right)^\top\right]$. Taking the derivative with respect to the components of $\mathbf{b}\left(\mathbf{u}_t\right)$ and setting the result equal to zero yields 
$\mathbf{b}\left(\mathbf{u}_t\right)=-\boldsymbol\Omega_t^{-1}\boldsymbol\omega_t$. 
The second derivative w.r.t. $\mathbf{b}$ corresponds to $\boldsymbol\Omega$ and is positive definite as long as the components of $\boldsymbol\psi$ are linearly independent. We have therefore found a unique minimum.

Assume now that $\mathbf{u}$ follows the dynamics $d\mathbf{u}_t=\boldsymbol\Delta_t\boldsymbol\Sigma_td\mathbf{w}_t$ and, for some $r<s$, consider $\boldsymbol\omega_r=\mathbbm{E}_r\left[\left(\mathbbm{E}_t\left[a\left(\mathbf{u}_T\right)\right]-\mathbbm{E}_r\left[a\left(\mathbf{u}_T\right)\right]\right)\left(\mathbf{u}_s-\mathbf{u}_r\right)\right]$ and $\boldsymbol\Omega_r=\mathbbm{E}_r\left[\left(\mathbf{u}_s-\mathbf{u}_r\right)\left(\mathbf{u}_s-\mathbf{u}_r\right)^\top\right]$. Then, as $r\rightarrow s$ (continuous monitoring), we have $\boldsymbol\Omega_t=d\left\langle\mathbf{u}\right\rangle_t$ and $\boldsymbol\omega_t=d\left\langle\mathbf{u}\right\rangle_t\mathbf{J}^a\left(\mathbf{u}_t\right)$ (using that $d\mathbbm{E}_t\left[a\left(\mathbf{u}_T\right)\right]=\mathbf{J}^a\left(\mathbf{u}_t\right)^\top d\mathbf{u}_t$), and therefore $\mathbf{b}\left(\mathbf{u}\right)=-\boldsymbol\Omega^{-1}\boldsymbol\omega=-\mathbf{J}^a\left(\mathbf{u}\right)$.

Alternatively, the same result can be obtained from first-order Taylor expansion. To see this, consider $f\left(\mathbf{u},\mathbf{u}+\delta\mathbf{u}\right)=\left\{\mathbf{J}^a\left(\mathbf{u}\right)+\mathbf{b}\left(\mathbf{u}\right)\right\}^\top\delta\mathbf{u}+O\left(\left|\delta\mathbf{u}\right|^2\right)$, and note that the leading order of $\delta\mathbf{u}$ is zero for $\mathbf{b}\left(\mathbf{u}\right)=-\mathbf{J}^a\left(\mathbf{u}\right)$.\qed\\

\noindent {\bfseries Proof of Corollary 4:} Inserting the specifications of $\mathbf{u}$, $a$ and $\mathbf{b}$ in \eqref{eq:msoln} yields
\begin{eqnarray*}
	f\left(\mathbf{u}_r,\mathbf{u}_s\right)&=&\left(c_5+2c_6^\lambda-2c_6^\eta\right)\left(\ln F_s-\ln F_r\right)+4c_8\left(Y_s^2-Y_r^2\right)+2c_6^\eta\left(\tfrac{Z_s}{F_s}-\tfrac{Z_r}{F_r}\right)\\
	&&+\left(\tfrac{c_7}{F_r}-\tfrac{2c_9Z_r}{F_r^2}\right)\left(F_s-F_r\right)+\left(-2c_6^\lambda-8c_8Y_r\right)\left(Y_s-Y_r\right)+\tfrac{2c_9}{F_r}\left(Z_s-Z_r\right)\\
	&=&c_5\left(y_s-y_r\right)+2c_6^\lambda\left(y_s-y_r-Y_s+Y_r\right)+2c_6^\eta\left(\tfrac{Z_s}{F_s}-\tfrac{Z_r}{F_r}-y_s+y_r\right)\\
	&&+c_7\mathrm{e}^{-y_r}\left(\mathrm{e}^{y_s}-\mathrm{e}^{y_r}\right)+4c_8\left(Y_s^2-2Y_r\left(Y_s-Y_r\right)-Y_r^2\right)\\
	&&+\tfrac{2c_9}{F_r}\left(Z_s-Z_r-\tfrac{Z_r}{F_r}\left(F_s-F_r\right)\right)\\
	&=&c_5\left(y_s-y_r\right)+c_6^\lambda\left(v_s^\lambda-v_r^\lambda\right)+c_6^\eta\left(v_s^\eta-v_r^\eta\right)+c_7\left(\mathrm{e}^{y_s-y_r}-1\right)\\
	&&+c_8\left(2Y_s-2Y_r\right)^2+c_9\left(\tfrac{2Z_s}{F_s}-\tfrac{2Z_r}{F_r}\right)\mathrm{e}^{y_s-y_r}=g\left(\mathbf{x}_s-\mathbf{x}_r\right).\qquad \qed\\
\end{eqnarray*}

\end{document}